\documentclass[
  aps,
  prd,
  reprint,
  nofootinbib,
  superscriptaddress,
  amsmath,amssymb,floatfix
]{revtex4-2}
\usepackage{booktabs}
\usepackage{multirow}
\usepackage{graphicx}
\usepackage{dcolumn}
\usepackage{bm}
\usepackage{hyperref}

\begin{document}

\preprint{APS/123-QED}

\title{\boldmath Redshift-binned constraints on the Hubble constant under $\Lambda$CDM, CPL, and Pad\'e cosmography}

\author{Zhi-Yuan Mo}
\affiliation{Institute for Frontiers in Astronomy and Astrophysics, 
Beijing Normal University, Beijing 102206, People’s Republic of China}
\affiliation{School of Physics and Astronomy, 
Beijing Normal University, Beijing 100875, People’s Republic of China}

\author{Kang Jiao}
\email{kangjiao@zzu.edu.cn}
\thanks{Corresponding author}
\affiliation{Institute for Astrophysics, School of Physics, 
Zhengzhou University, Zhengzhou 450001, People’s Republic of China}

\author{Tong-Jie Zhang}
\email{tjzhang@bnu.edu.cn}
\thanks{Corresponding author}
\affiliation{Institute for Frontiers in Astronomy and Astrophysics, 
Beijing Normal University, Beijing 102206, People’s Republic of China}
\affiliation{School of Physics and Astronomy, 
Beijing Normal University, Beijing 100875, People’s Republic of China}

\date{Received 9 December 2025; accepted 21 January 2026; published 13 February 2026}

\begin{abstract}
Motivated by recent claims of a possible redshift dependence in late-Universe determinations of the Hubble constant ($H_0$), we test the robustness of this behavior using multiple cosmological probes. We perform a joint redshift-binned analysis of $H_0$ across eight bins using late-Universe probes—Pantheon+ type Ia supernova, Dark Energy Spectroscopic Instrument baryon acoustic oscillation, cosmic chronometers, and water megamasers—under three cosmological frameworks: flat $\Lambda$ cold dark matter, Chevallier-Polarski-Linder, and Pad\'e cosmography.
Under a common baseline scheme, all three models show a qualitatively similar, low-amplitude variation in the per-bin $H_0$ estimates. A simple Fourier-like parametrization captures this behavior, but the amplitude differs from zero only at a marginal significance of about $1.71$--$1.94\,\sigma$, with similar behavior observed across all three cosmological frameworks.
We then investigate the robustness and possible origin of this feature. Alternative binning schemes preserve its qualitative form, whereas single-probe per-bin fits (supernova-only, cosmic chronometer-only, BAO-only) yield ratios $H_{0,i}/H_{0,\mathrm{global}}$ mostly consistent with unity and do not reproduce the pronounced drift seen in the joint baseline constraints. Finally, by comparing different global versus piecewise-constant configurations for $\{H_0,\Omega_m,M,r_d\}$, we find that a baselinelike oscillatory pattern reemerges only when multiple degenerate parameter combinations are allowed to vary across bins, while it is strongly suppressed when only $H_0$ is bin dependent.
Taken together, these results indicate that the apparent oscillatory behavior of $H_0(z)$ in late time arises from known parameter degeneracies and does not constitute robust evidence for a genuine redshift evolution.
\end{abstract}

\keywords{Hubble tension, Cosmography, Bayesian analysis, $H_0(z)$ evolution}
\maketitle


\section{Introduction}
Since the late 1990s, the $\Lambda$ cold dark matter ($\Lambda$CDM) model has provided the standard framework of modern cosmology, with only a few basic parameters giving a clear and useful description of the Universe that matches a wide range of observations, including the cosmic microwave background (CMB) anisotropies~\cite{Planck2018}, baryon acoustic oscillations (BAO) from galaxy surveys~\cite{Eisenstein2005,Alam2017}, the clustering of large-scale structure~\cite{Alam2017}, and type~Ia supernovae (SNe~Ia)~\cite{Suzuki2012,Betoule2014,Brout2022}. Despite this overall success, high-precision measurements of the Hubble constant $H_0$ from early- and late-Universe probes exhibit a persistent discrepancy at the $\sim 5\sigma$ level, commonly referred to as the \emph{Hubble tension}. In particular, the value of $H_0$ inferred from \textit{Planck} CMB data within $\Lambda$CDM is significantly lower than that obtained from local distance-ladder measurements calibrated by Cepheid variables and SNe~Ia~\cite{Planck2018,Riess2022}.

Within standard Friedmann-Lema\^itre-Robertson-Walker (FLRW) cosmology,, the Hubble constant $H_0$ is a single global parameter defined at $z=0$. However, several recent analyses indicate that the values of $H_0$ inferred at different redshifts are mutually inconsistent, hinting at redshift-dependent systematics or deviations from the standard $\Lambda$CDM framework.
Early hints of a redshift dependence in $H_0$ came from strong-lensing time-delay cosmography.  
Bonvin et al.~\cite{Bonvin2017} noted that the first three H0LiCOW systems suggested higher $H_0$ values at lower lens redshifts, although the small sample size precluded firm conclusions.  
With six lenses, Wong et al.~\cite{Wong2019} confirmed that this trend persists and reported a mild indication of decreasing $H_0$ with increasing redshift, in a way fully independent of both distance-ladder and CMB approaches.  
Subsequently, Krishnan et al.~\cite{Krishnan2020} performed a complementary analysis with a low-redshift ($z<0.7$) dataset comprising megamasers, cosmic chronometers, SNe~Ia, and BAO.  
By dividing the data into redshift bins and fitting the data in each bin with a flat $\Lambda$CDM model, they found a decreasing trend in $H_0$ with redshift, with the fitted slope significant at the $2.1\sigma$ level.
Following this, Dainotti et al.~\cite{Dainotti2024} applied a similar binning approach to the Pantheon SNe~Ia sample and also reported a decreasing tendency of $H_0$.  
Lopez-Hernandez and De-Santiago~\cite{LopezHernandez2025} extended the analysis to a wider redshift range ($0<z<2.3$) with multiple probes, finding indications of a dynamical evolution of $H_0$. Within the $w_0w_a$CDM framework, they reported consistent results, with the statistical significance of the trend depending on the parameterization, lying between $1.5\sigma$ and $2.3\sigma$.  
More recently, Dainotti et al.~\cite{Dainotti2025} combined several SNe~Ia compilations into a unified master dataset and, using multiple binning strategies with Markov Chain Monte Carlo (MCMC) analyses, showed that the decreasing trend of $H_0$ can be phenomenologically characterized by $\alpha \sim 0.01$.

While these analyses report possible hints of a redshift-dependent \(H_0(z)\), they have rarely explored, within a single unified framework, which aspects of the methodology drive the observed trends. Our analysis fills this gap by sequentially examining the effects of cosmological parametrization, binning strategy, probe combination, and parameter-sharing assumptions, thereby clarifying how each of these ingredients --- and their interplay --- can shape the apparent evolution of \(H_0(z)\).

We revisit this question within a unified, redshift-binned, multiprobe and multimodel framework, motivated by the recent analysis of Ref.~\cite{LopezHernandez2025}, which reported an oscillatory pattern in binned late-time determinations of $H_0$. 
We combine Pantheon+ SNe~Ia, Dark Energy Spectroscopic Instrument (DESI) BAO, cosmic chronometers, and water megamasers, and perform a per-bin analysis of $H_0$ under three cosmological frameworks: flat $\Lambda$CDM, Chevallier-Polarski-Linder, and Pad\'e cosmography.
Using a common eight-bin baseline scheme, we first obtain per-bin constraints on $H_0$ and quantify any apparent oscillatory behavior through a simple Fourier-like parameterization. 
We then investigate the robustness and possible origin of the inferred pattern by examining three aspects: 
(i) alternative binning choices, 
(ii) single-probe per-bin fits, and 
(iii) different global versus piecewise-constant parameter configurations for $\{H_0,\Omega_m,M,r_d\}$, which allow us to assess the role of probe-specific parameter degeneracies.

The structure of this paper is as follows. Section~\ref{sec:data_method} describes the datasets, the adopted binning schemes, the cosmological models, and the statistical methodology. Section~\ref{sec:results} presents the baseline redshift-binned constraints on \(H_0\) and their Fourier-like characterization. Section~\ref{sec:Analysis} examines the robustness and interpretation of these results through alternative binning schemes, single-probe fits, and global versus piecewise-constant parameter configurations. Finally, Sec.~\ref{sec:conclusions} summarizes our conclusions and discusses their implications.

\section{Datasets and Methodology}
\label{sec:data_method}

\subsection{Datasets}
\label{sec:data}
The datasets used in this work are consistent with those adopted in Ref.~\cite{LopezHernandez2025} and include four complementary late-time cosmological probes.

\paragraph{SNe~Ia.}
We use the Pantheon+ dataset, consisting of 1701 light curves of 1550 unique, spectroscopically confirmed SNe~Ia covering the redshift range $0.01 \leq z \leq 2.26$~\cite{Scolnic2022,Brout2022}.
Pantheon+ provides standard candles that are crucial for constraining $M$ and the relative distance scale.
To avoid introducing late-Universe distance-ladder information through externally calibrated distances, we exclude the 77 SNe in Cepheid host galaxies that form the Pantheon+ calibration subsample, following Ref.~\cite{Malekjani2024}.
The absolute magnitude $M$ is then treated as a free parameter in the fit.

\paragraph{BAO.}
We employ DESI Data Release 2 (DR2) BAO measurements in the range $0.30 \leq z \leq 2.33$, including both isotropic ($D_V/r_d$) and anisotropic ($D_M/r_d$, $D_H/r_d$) observables, which measure the comoving distance $D_M(z)$ and the expansion rate $H(z)$.
To avoid introducing early-Universe information from the CMB, we treat the sound-horizon scale $r_d$ as a free parameter in our fits.
The covariance matrices are reconstructed from the published uncertainties and correlation coefficients~\cite{DESI_DR2}.

\paragraph{Cosmic chronometers (CC).}
We use 37 independent $H(z)$ determinations spanning $0.07 \leq z \leq 1.965$, 
which are summarized in Table~\ref{tab:cc_data}. 
The CC data yield independent constraints on the cosmological parameters and help to reduce parameter degeneracies, for example between $H_0$ and $M$ or $r_d$.
The covariance matrix is estimated following the prescription of Moresco.\footnote{\url{https://gitlab.com/mmoresco/CCcovariance}}
For the data point at $z=0.09$, the commonly quoted value $H(0.09)=69\pm12~\mathrm{km\,s^{-1}\,Mpc^{-1}}$ originates from Ref.~\cite{Jimenez2003}, where it was used as an estimate of the Hubble constant $H_0$, rather than a direct measurement of $H(z)$. Following the correction procedure of Ref.~\cite{Niu2025}, we convert this $H_0$ value into $H(z=0.09)$ using the cosmological parameters adopted in that work, yielding $H(0.09)=70.7\pm12.3~\mathrm{km\,s^{-1}\,Mpc^{-1}}$.

\paragraph{Megamasers.}
We include distances and recession velocities for six water–megamaser host galaxies within $0.002 \leq z \leq 0.034$, provided by the Megamaser Cosmology Project~\cite{Pesce2020}. 
Megamasers serve as a local low-redshift anchor on the Hubble constant when combined with the SNe~Ia sample.

\begin{table}[t]
\caption{List of the 37 cosmic-chronometer $H(z)$ measurements used in this work.}
\label{tab:cc_data}
\begin{ruledtabular}
\begin{tabular}{cccccccc}
$z$ & $H(z)$ & $\sigma_H$ & Ref. & $z$ & $H(z)$ & $\sigma_H$ & Ref. \\
\hline
0.07  & 69.0   & 19.6   & \cite{Zhang2014}     & 0.5   & 72.1   & 34.68  & \cite{Loubser2025} \\
0.09  & 70.7   & 12.3   & \cite{Jimenez2003}     & 0.593 & 104.0  & 13.0   & \cite{Moresco2012} \\
0.12  & 68.6   & 26.2   & \cite{Zhang2014}     & 0.67  & 119.45 & 17.82  & \cite{Loubser2025CC} \\ 
0.17  & 83.0   & 8.0    & \cite{Simon2005}     & 0.68  & 92.0   & 8.0    & \cite{Moresco2012} \\
0.179 & 75.0   & 4.0    & \cite{Moresco2012}   & 0.781 & 105.0  & 12.0   & \cite{Moresco2012} \\
0.199 & 75.0   & 5.0    & \cite{Moresco2012}   & 0.8   & 113.1  & 25.22  & \cite{Jiao2023} \\
0.2   & 72.9   & 29.6   & \cite{Zhang2014}     & 0.83  & 108.28 & 18.13  & \cite{Loubser2025CC} \\ 
0.27  & 77.0   & 14.0   & \cite{Simon2005}     & 0.875 & 125.0  & 17.0   & \cite{Moresco2012} \\
0.28  & 88.8   & 36.6   & \cite{Zhang2014}     & 0.88  & 90.0   & 40.0   & \cite{Stern2010} \\
0.352 & 83.0   & 14.0   & \cite{Moresco2012}   & 0.9   & 117.0  & 23.0   & \cite{Simon2005} \\
0.38  & 83.0   & 13.5   & \cite{Moresco2016}   & 1.037 & 154.0  & 20.0   & \cite{Moresco2012} \\
0.4   & 95.0   & 17.0   & \cite{Simon2005}     & 1.26  & 135.0  & 65.0   & \cite{Tomasetti2023} \\
0.4004 & 77.0   & 10.2  & \cite{Moresco2016}   & 1.3   & 168.0  & 17.0   & \cite{Simon2005} \\
0.425 & 87.1   & 11.2   & \cite{Moresco2016}   & 1.363 & 160.0  & 33.6   & \cite{Moresco2015} \\
0.445 & 92.8   & 12.9   & \cite{Moresco2016}   & 1.43  & 177.0  & 18.0   & \cite{Simon2005} \\
0.46  & 88.48  & 12.33  & \cite{Loubser2025CC}                & 1.53  & 140.0  & 14.0   & \cite{Simon2005} \\ 
0.47  & 89.0   & 49.6   & \cite{Ratsimbazafy2017} & 1.75  & 202.0  & 40.0   & \cite{Simon2005} \\
0.4783 & 80.9  & 9.0    & \cite{Moresco2016}   & 1.965 & 186.5  & 50.4   & \cite{Moresco2015} \\
0.48  & 97.0   & 62.0   & \cite{Stern2010}     &       &        &        &  \\
\end{tabular}
\end{ruledtabular}
\end{table}

\subsection{Binning strategy}
\label{sec:A0_binning}

We divide these datasets into eight redshift intervals following the framework introduced by Ref.~\cite{LopezHernandez2025}. 
This binning strategy is guided by two requirements: each bin must retain sufficient constraining power and admit a well-defined effective redshift. 
We therefore choose the boundaries such that the effective redshifts of different probes within the same interval are mutually consistent, so that multiprobe measurements assigned to a given bin can be interpreted as probing the same epoch of cosmic expansion~\cite{Krishnan2020}.

For each bin $i$, the effective redshift $\bar{z}_i$ is defined as an inverse--variance--weighted mean:
\begin{equation}
\bar{z}_i = \frac{\sum_{k=1}^{N_i} z_k\,\sigma_k^{-2}}{\sum_{k=1}^{N_i} \sigma_k^{-2}}\,,
\label{eq:z_eff}
\end{equation}
where $N_i$ is the number of measurements in the bin and $\sigma_k$ the $1\sigma$ uncertainty of the corresponding observable.  

The intermediate- and high-redshift bins ($0.30 \lesssim z \lesssim 2.33$) are defined primarily by the distribution of DESI BAO measurements. 
In practice, bins~3--8 are chosen such that each contains at least one BAO data point and the effective redshift $z_{\mathrm{eff}}$ of each bin closely matches the redshift of the corresponding BAO observation.  

At low redshift, the lower boundary of bin~2 is set by the minimum redshift of the cosmic-chronometer sample, ensuring that all CC measurements lie at or above this boundary.
The lowest-redshift measurements, including all water megamasers and the corresponding nearby SNe~Ia, are assigned to bin~1. 
This data-driven construction ensures that the bin boundaries reflect the actual redshift distribution of the combined datasets, rather than an arbitrary segmentation. 
The detailed bin definitions and effective redshifts are summarized in Table~\ref{tab:binningA0}.

This configuration serves as the primary binning strategy for all main analyses in this work. 
We refer to this configuration as the primary binning scheme (A0).
To further examine whether the observed $H_0(z)$ trend arises from this particular binning choice,  
we also test three alternative binning schemes (A1–A3) in Sec.~\ref{sec:alt_binning}.

\begin{table}[t]
\caption{Primary binning scheme (A0): summary of the redshift bins used in this work.
The numbers in parentheses indicate the number of data points from each dataset included in each bin.}
\label{tab:binningA0}
\begin{ruledtabular}
\begin{tabular}{clcc}
Bin & Data sets included & Redshift range & $z_{\rm eff}$ \\
\hline\noalign{\vskip 0.3ex}
1 & SNe (572), Megamasers (6) & $(0.01,\,0.069]$ & 0.032 \\
2 & SNe (250), CC (6) & $(0.069,\,0.199]$ & 0.150 \\
3 & SNe (482), CC (8), BAO (1) & $(0.199,\,0.425]$ & 0.300 \\
4 & SNe (162), CC (7), BAO (2) & $(0.425,\,0.625]$ & 0.510 \\
5 & SNe (79), CC (3), BAO (2) & $(0.625,\,0.7891]$ & 0.710 \\
6 & SNe (14), CC (6), BAO (2) & $(0.7891,\,1.13]$ & 0.930 \\
7 & SNe (15), CC (5), BAO (4) & $(1.13,\,1.65]$ & 1.380 \\
8 & SNe (4), CC (2), BAO (2) & $(1.65,\,2.33]$ & 1.940 \\
\end{tabular}
\end{ruledtabular}
\end{table}

\subsection{Cosmological models}
\label{sec:cosmo_models}
Using the redshift-binning strategy described above, we derive constraints on $H_0$ within three cosmological frameworks: flat $\Lambda$CDM, the CPL parametrization of dark energy, and Pad\'e cosmography.
This multimodel comparison allows us to distinguish effects driven by parametric assumptions from those that are robustly supported by the data.

\paragraph{Flat $\Lambda$CDM.}
As our baseline, we assume a spatially flat Universe with pressureless matter and a cosmological constant~\cite{Peebles1993}. The Hubble rate is
\begin{equation}
H(z) \;=\; H_0\,\sqrt{\Omega_m(1+z)^3 + (1-\Omega_m)}\,,
\label{eq:Hz_LCDM}
\end{equation}
where $\Omega_m$ is the matter density parameter today. We neglect radiation, since it is subdominant over the redshift range considered ($z\lesssim 2.3$). Distance measures follow from the standard relations; for example, the luminosity distance is
\begin{equation}
d_L(z) \;=\; (1+z)\,c \int_0^z \frac{dz'}{H(z')}\, .
\label{eq:dL_general}
\end{equation}
This model is characterized by the parameter vector $\{ H_{0}, \Omega_{m} \}$.

\paragraph{Chevallier--Polarski--Linder (CPL) parametrization.}
To allow for a redshift-dependent dark-energy equation of state, we adopt the CPL form~\cite{Chevallier2001,Linder2003},
\begin{equation}
w(z) \;=\; w_0 + w_a\,\frac{z}{1+z}\, .
\end{equation}
In a flat Universe, the corresponding Hubble rate can be written as
\begin{equation}
\label{eq:Hz_CPL}
\begin{aligned}
H(z) &= H_0 \sqrt{\,\Omega_m(1+z)^3 + (1-\Omega_m)\,f(z;w_0,w_a)\,},\\
f(z;w_0,w_a) &= (1+z)^{3(1+w_0+w_a)}\,
\exp\!\left[-\frac{3 w_a z}{1+z}\right].
\end{aligned}
\end{equation}
The CPL parametrization reduces to $\Lambda$CDM for $(w_0,w_a)=(-1,0)$. 
It thus provides a controlled extension of the baseline that is flexible enough to capture smooth late-time dynamics while introducing only two additional parameters.

\paragraph{Pad\'e cosmography.}
We also consider a Pad\'e parametrization, which we use as a purely kinematic description constructed directly from the FLRW metric without assuming a specific dark-energy equation of state~\cite{Visser2005,Gruber2014,Zhou2016,Capozziello2019a,Capozziello2019b,Aviles2014,Hu2024}. 
This provides a flexible phenomenological baseline that we analyze on the same footing as the dynamical models and use as a cross-check of model dependence.

In standard cosmography, cosmological observables such as the Hubble rate $H(z)$ and the luminosity distance $d_L(z)$ are typically expanded in Taylor series around $z=0$. 
However, Taylor expansions suffer from poor convergence at higher redshifts, which can bias the estimation of cosmological parameters when applied to real data extending beyond $z\gtrsim 1$~\cite{Pourojaghi2025}. 
To address this limitation we instead use Pad\'e approximants. 
A Pad\'e approximant $P_{n\,m}(z)$ is a rational function
\begin{equation}
P_{n\,m}(z) \;=\; \frac{a_0 + a_1 z + \dots + a_n z^n}{1 + b_1 z + \dots + b_m z^m}\,,
\end{equation}
and this rational structure typically yields improved convergence over comparable Taylor truncations.

As discussed in Ref.~\cite{Petreca2024}, Pad\'e cosmography is a flexible tool, but we must choose the orders $(n,m)$ with care, given the redshift span and the number of kinematical parameters we aim to constrain.
In this work we adopt the $P_{21}$ form, which strikes a good balance between stability and parsimony in the late-time regime. 
Higher-order choices (e.g.\ $P_{32}$) can incorporate additional parameters such as the snap $s_0$, but $P_{21}$ already delivers competitive performance with fewer degrees of freedom.

Following the approach of Ref.~\cite{Aviles2014}, we apply the Pad\'e expansion directly to the luminosity distance $d_L(z)$ rather than to $H(z)$, which helps control the high-$z$ behavior. 
Specifically, we use the $P_{21}$ approximant
\begin{equation}
d_L^{P_{21}}(z) \;=\; \frac{c z}{H_0}\,
\frac{6(q_0 - 1) + z\!\left[-5 - 2j_0 + q_0(8 + 3q_0)\right]}
{-2(3 + z + j_0 z) + 2q_0(3 + z + 3q_0 z)}\,,
\label{eq:d_L}
\end{equation}
which is expressed in terms of the cosmographic set $\{H_0, q_0, j_0\}$~\cite{visser2004,Aviles2014}. 
When needed (e.g.\ to predict CC/BAO observables), we obtain the expansion rate from this distance representation via the flat-geometry identity
\begin{equation}
\frac{c}{H(z)} \;=\; \frac{d}{dz}\!\left[\frac{d_L(z)}{1+z}\right].
\label{eq:Hz_from_dL}
\end{equation}
Pad\'e cosmography can also be reformulated in terms of dark-energy parametrizations through an appropriate parameter mapping~\cite{Wang2009}. 
In our analysis the corresponding parameter vector is $\{ H_{0}, q_{0}, j_{0} \}$.

\subsection{MCMC setup and likelihoods}

To estimate the parameter vector $\boldsymbol{\Theta}$ in each redshift bin, we perform a Bayesian MCMC analysis using the \texttt{EMCEE} \texttt{PYTHON} package~\cite{ForemanMackey2013}. This sampler implements the affine-invariant ensemble algorithm, which is well suited for exploring multidimensional likelihood surfaces.

We adopt a common MCMC setup for all models and redshift bins. Specifically, we use 48 chains, each with 4000 steps, and 1000 burn-in steps. This ensures that each model is sampled thoroughly within each redshift bin. We monitor convergence using the Gelman-Rubin diagnostic.

\subsubsection{Priors}
We adopt flat (uniform) priors for all free parameters, reflecting our aim not to favor any particular region of parameter space \textit{a priori}.
The prior limits for each parameter are summarized in Table~\ref{tab:priors}. 
These ranges are chosen to be broad enough to explore the parameter space comprehensively,
while remaining consistent with physical considerations and previous studies~\cite{Krishnan2020, Aviles2014, LopezHernandez2025}.

\begin{table}[t]
\centering
\caption{Flat prior ranges adopted for all free parameters.
Units are $H_0$ in km\,s$^{-1}$\,Mpc$^{-1}$ and $r_d$ in Mpc.}
\label{tab:priors}
\begin{ruledtabular}
\begin{tabular}{lcc}
\textbf{Parameter} & \textbf{Description} & \textbf{Prior range} \\
\hline
$H_0$      & Hubble constant                & $0 < H_0 < 100$ \\
$\Omega_m$ & Matter density parameter       & $0 < \Omega_m < 1$ \\
$M$        & SN absolute magnitude          & $-50 < M < 0$ \\
$r_d$      & Sound horizon                  & $0 < r_d < 200$ \\
$w_0$      & Dark energy (DE) equation & $-1.5 < w_0 < -1/3$ \\
           & of state (EOS) $z = 0$ & \\
$w_a$      & DE evolution parameter         & $-2.5 < w_a < 2.5$ \\
$q_0$      & Deceleration parameter         & $-3 < q_0 < 1.5$ \\
$j_0$      & Jerk parameter                 & $-3 < j_0 < 3$ \\
\end{tabular}
\end{ruledtabular}
\end{table}

\subsubsection{Likelihood}
\label{sec:likelihood}
Having specified the priors, we now define the likelihood function used to evaluate the model parameters. 
The likelihood takes the standard Gaussian form
\begin{equation}
\mathcal{L} \;\propto\; \exp\!\left(-\tfrac{1}{2}\chi^2\right),
\end{equation}
where the chi-squared statistic $\chi^2$ is constructed according to the specific 
observables of each dataset, as described in the following.

\paragraph{Type Ia supernovae.}
For the Pantheon+ sample, the observable is the corrected apparent magnitude 
$m_{\text{obs}}$, and the theoretical prediction is
\begin{equation}
m_{\text{th}}(z) = 5 \log_{10} \left[ \frac{d_L(z)}{\mathrm{Mpc}} \right] + M +25 \,,
\end{equation}
where $M$ denotes the absolute-magnitude calibration parameter. The chi-squared 
function is defined as~\cite{Brout2022}
\begin{equation}
\chi^2_{\rm SN} = (\mathbf{m}_{\rm obs} - \mathbf{m}_{\rm th})^{\rm T} 
\mathbf{C}^{-1} 
(\mathbf{m}_{\rm obs} - \mathbf{m}_{\rm th}) \,,
\end{equation}
with $\mathbf{C}$ the full statistical and systematic covariance matrix.

\paragraph{DESI-BAO.}
BAO observables are commonly expressed through characteristic distance measures along and across the line of sight.  
Along the line of sight, the Hubble distance is defined as
\begin{equation}
D_H(z) = \frac{c}{H(z)}\,,
\label{eq:DH}
\end{equation}
where \( c = 299{,}792.458\,\text{km/s} \) is the speed of light.  
In the transverse direction, the comoving angular diameter distance is
\begin{equation}
D_M(z) = (1+z) D_A(z)\,,
\label{eq:DM1}
\end{equation}
with \( D_A(z) \) the angular diameter distance.  
The angular diameter distance is given by
\begin{equation}
D_A(z) = \frac{c}{1+z} \int_0^z \frac{dz'}{H(z')}\,.
\label{eq:DA}
\end{equation}
The isotropic angle-averaged distance, relevant for spherically averaged BAO, is then
\begin{equation}
D_V(z) = \left[ z \cdot D_M^2(z) \cdot D_H(z) \right]^{1/3}\,.
\label{eq:DV}
\end{equation}

Depending on the type of BAO data available in each redshift bin, we consider two likelihood forms:

\textit{(i) Isotropic BAO.}  
For bins containing isotropic BAO measurements, the chi-squared function is
\begin{equation}
\chi^2_{\text{iso}} = \sum_{i=1}^{n} \frac{ \left[ \left( D_V / r_d \right)^{\text{th}} (z_i) - \left( D_V / r_d \right)^{\text{obs}}(z_i) \right]^2 }{ \sigma_i^2 }\,.
\label{eq:chi2_iso}
\end{equation}

\textit{(ii) Anisotropic BAO.}  
For bins containing anisotropic BAO measurements, we adopt the correlated chi-squared form
\begin{equation}
\chi^2_{\text{aniso}} = \mathbf{Q}^\top \cdot \mathbf{C}_{\text{BAO}}^{-1} \cdot \mathbf{Q}\,,
\label{eq:chi2_aniso}
\end{equation}
where the residual vector \( \mathbf{Q} \) is defined by
\begin{equation}
Q_i =
\begin{cases}
\bigl(D_M / r_d\bigr)^{\text{th}}(z_i)
 - \bigl(D_M / r_d\bigr)^{\text{obs}}(z_i),\\[2pt]
\bigl(D_H / r_d\bigr)^{\text{th}}(z_i)
 - \bigl(D_H / r_d\bigr)^{\text{obs}}(z_i),
\end{cases}
\label{eq:Qi}
\end{equation}
where the upper (lower) line applies to $D_M$ ($D_H$) measurements.
This chi-squared construction follows the treatment of Ref.~\cite{Pourojaghi2025}.

\paragraph{Cosmic chronometers.}
The CC dataset provides direct measurements of the Hubble parameter $H(z)$. 
The chi-squared function is defined as
\begin{equation}
\chi^2_{\text{CC}} = 
\left( \mathbf{H}_{\text{obs}} - \mathbf{H}_{\text{th}} \right)^{\mathrm{T}} 
\, \mathbf{C}_{\text{CC}}^{-1} \, 
\left( \mathbf{H}_{\text{obs}} - \mathbf{H}_{\text{th}} \right)\,,
\label{eq:chi2_cc}
\end{equation}
where $\mathbf{H}_{\text{obs}}$ denotes the vector of observed values and 
$\mathbf{H}_{\text{th}}$ the corresponding model predictions.

\paragraph{Megamasers.}
Following Ref.~\cite{Pesce2020}, we convert between velocities and redshift using the relation \(v = cz\).  
To account for uncertainties in the peculiar velocities, we inflate the error by \(\sigma_{\text{pec}} = 250 \, \text{km/s}\).  
The chi-squared function is then given by
\begin{equation}
\chi^2_{\text{maser}} = \sum_{i=1}^{6} \left[ \frac{(v_i - \hat{v}_i)^2}{\sigma_{v,i}^2 + \sigma_{\text{pec}}^2} + \frac{(D_A(z_i) - \hat{D}_i)^2}{\sigma_{D,i}^2} \right]\,,
\label{eq:chi2_maser}
\end{equation}
where \(v_i\) are the observed velocities treated as nuisance parameters, and \(\hat{v}_i, \hat{D}_i, \sigma_{v,i}, \sigma_{D,i}\)
denote the inferred velocities, galaxy distances, and their corresponding uncertainties from modeling maser disks~\cite{Pesce2020}.

\section{Results}
\label{sec:results}

\subsection{Baseline scheme results}
\label{sec:baseline}

We first present the baseline constraints on the Hubble constant $H_0$ obtained by combining the four late-time probes (Pantheon+, DESI DR2 BAO, CC, and water megamasers) under the primary eight-bin scheme A0 described in Sec.~\ref{sec:A0_binning}. 
For each bin we independently fit the three cosmological frameworks introduced in Sec.~\ref{sec:cosmo_models} (flat $\Lambda$CDM, CPL, and Pad\'e cosmography) using the common MCMC setup of Sec.~\ref{sec:likelihood}, and extract per-bin posteriors for $H_0$.

\begin{figure}[t]
  \centering
  \includegraphics[width=\columnwidth]{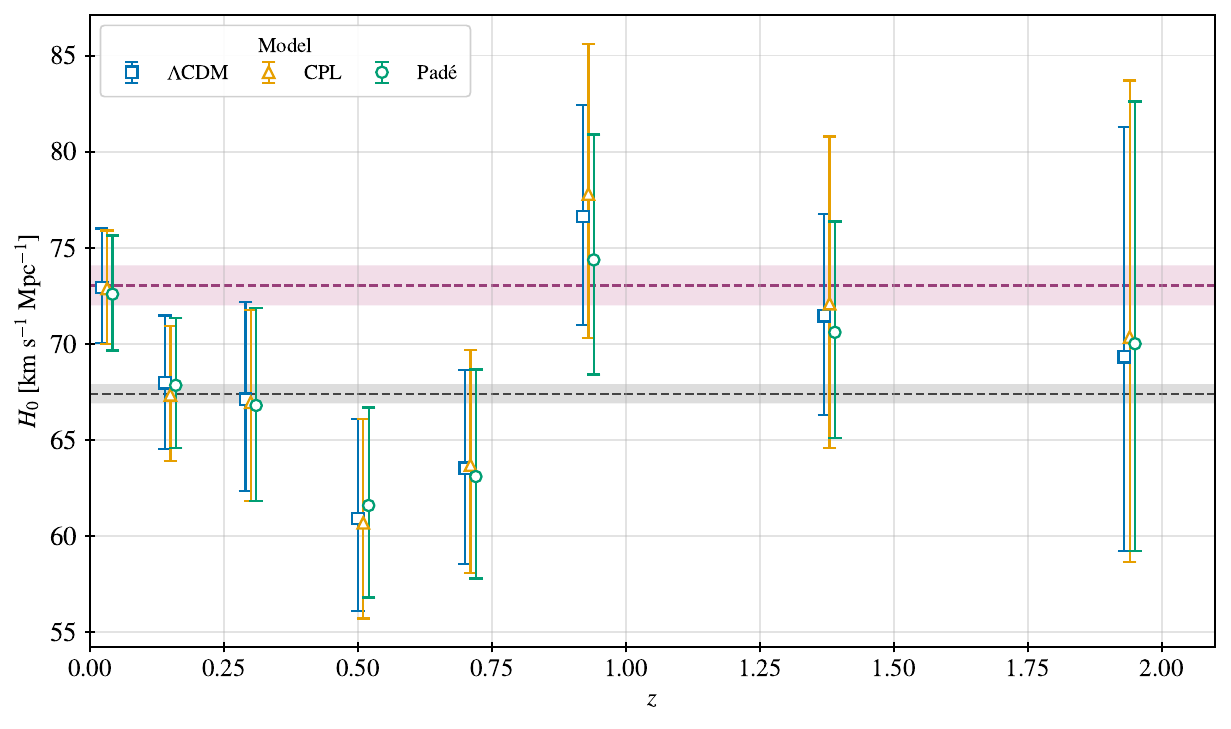}
  \caption{Per-bin constraints on the Hubble constant $H_0(z)$ from three models under the baseline scheme.
Blue squares: flat $\Lambda$CDM; orange triangles: CPL ($w_0w_a$CDM); green circles: Pad\'e $P_{21}$ cosmography.
Error bars show 68\% credible intervals at the effective redshift $z_{\rm eff}$.
Purple and gray shaded bands indicate the \textit{SH0ES}
($H_0 = 73.04 \pm 1.04$~km\,s$^{-1}$\,Mpc$^{-1}$)~\cite{Riess2022}
and \textit{Planck} 2018
($H_0 = 67.4 \pm 0.5$~km\,s$^{-1}$\,Mpc$^{-1}$)~\cite{Planck2018} results, respectively.}
  \label{fig:baseline}
\end{figure}

Figure~\ref{fig:baseline} shows the posterior medians and $68\%$ credible intervals of $H_0$ in each of the eight redshift bins for the three models (points placed at the effective redshifts $z_{\rm eff}$ from Table~\ref{tab:binningA0}). 
Horizontal shaded bands indicate the external anchors from \textit{Planck}~2018 and SH0ES~2022 for visual reference~\cite{Planck2018, Riess2022}; numerical values for the per-bin constraints are listed in Table~\ref{tab:h0_results}.

All three models reveal a qualitatively similar redshift dependence. 
Considering the joint constraints from the baseline scheme, $H_0$ decreases across the first four bins ($0 < z \lesssim 0.6$), then rises again in the intermediate range ($0.6 \lesssim z \lesssim 1.0$), and in the last two bins settles between the SH0ES~2022 and \textit{Planck}~2018 reference values. 
The three frameworks agree very well in the first five bins ($z \lesssim 0.8$), yielding nearly identical $H_0$ constraints and a common trend. 
In the last three, higher-redshift bins, small differences appear, with the largest model-to-model discrepancy being $3.43~\mathrm{km\,s^{-1}\,Mpc^{-1}}$, well within the corresponding error bars.
This indicates that the mild oscillatory-looking pattern in the binned $H_0(z)$ is primarily driven by the data and is not strongly dependent on the choice of background cosmological parametrization. 
In the next Section we quantify this behavior through a simple Fourier-like parametrization.

\begin{table}[t]
\centering
\caption{Hubble constant $H_0$ constraints in each redshift bin under the baseline scheme
for the flat $\Lambda$CDM, CPL, and Pad\'e models. Entries give the posterior median and
68\% credible interval in units of km\,s$^{-1}$\,Mpc$^{-1}$.}
\label{tab:h0_results}
\renewcommand{\arraystretch}{1.25}
\begin{ruledtabular}
\begin{tabular}{lccc}
Bin &
$\Lambda$CDM &
CPL &
Pad\'e \\
\hline\noalign{\vskip 0.3ex}
1 &
$72.94^{+3.07}_{-2.88}$ &
$72.88^{+3.01}_{-2.88}$ &
$72.59^{+3.06}_{-2.93}$ \\[1ex]
2 &
$67.97^{+3.51}_{-3.44}$ &
$67.35^{+3.59}_{-3.44}$ &
$67.84^{+3.52}_{-3.26}$ \\[1ex]
3 &
$67.12^{+5.05}_{-4.78}$ &
$66.96^{+4.80}_{-5.12}$ &
$66.80^{+5.08}_{-4.95}$ \\[1ex]
4 &
$60.91^{+5.18}_{-4.82}$ &
$60.69^{+5.41}_{-4.97}$ &
$61.59^{+5.10}_{-4.78}$ \\[1ex]
5 &
$63.53^{+5.11}_{-4.97}$ &
$63.70^{+5.98}_{-5.63}$ &
$63.10^{+5.58}_{-5.30}$ \\[1ex]
6 &
$76.63^{+5.79}_{-5.66}$ &
$77.80^{+7.82}_{-7.48}$ &
$74.37^{+6.52}_{-5.96}$ \\[1ex]
7 &
$71.46^{+5.29}_{-5.15}$ &
$72.09^{+8.70}_{-7.51}$ &
$70.60^{+5.77}_{-5.50}$ \\[1ex]
8 &
$69.32^{+11.98}_{-10.12}$ &
$70.36^{+13.35}_{-11.69}$ &
$70.01^{+12.61}_{-10.77}$ \\
\end{tabular}
\end{ruledtabular}
\end{table}

\subsection{Fourier parametrization}
\label{sec:fourier}

The baseline results in Sec.~\ref{sec:baseline} exhibit an apparent oscillatory pattern in the binned $H_0(z)$: a decrease over $z \simeq 0$--0.6, a rise around $z \sim 0.6$--1.0, and a mild decline in the two highest-redshift bins. 
To quantify this behavior in a model-agnostic way, we introduce a Fourier parametrization for $H_0(z)$ as a phenomenological probe of smooth, coherent departures from a constant $H_0$.

Our approach is motivated by Ref.~\cite{Tamayo2019}, which describes the dark-energy equation of state $w(z)$ in terms of a Fourier series. 
As emphasized there, trigonometric functions form a bounded, well-behaved basis over a finite redshift (or scale-factor) interval and can efficiently represent oscillatory patterns with only a few modes. 
In the same spirit, we adopt a general Fourier form for an effective Hubble constant,
\begin{equation}
H_0^{\mathrm{eff}}(z; \{b_k, c_k\})
=
\hat{H}_0
+
\sum_k
\bigl[
b_k \sin(2 \pi k a)
+
c_k \cos(2 \pi k a)
\bigr],
\label{eq:fourier_general}
\end{equation}
where \(a \equiv 1/(1+z)\).

Guided further by Ref.~\cite{LopezHernandez2025}, which found that a single cosine term already reproduces the oscillatory trend in binned $H_0$ constraints, we adopt the corresponding minimal two-parameter form
\begin{equation}
H_0^{\mathrm{eff}}(z)
=
\hat{H}_0
+
A \cos(4 \pi a) ,
\label{eq:fourier_singlemode}
\end{equation}
where $\hat{H}_0$ denotes the mean level and $A$ quantifies the amplitude of a smooth periodic deviation about this mean. 
This parametrization is purely phenomenological and is not tied to any specific underlying cosmological model; it is used solely as a fit to the redshift-dependent pattern encoded in the binned $H_0$ constraints.

The constant-$H_0$ case corresponds to the nested hypothesis $A=0$, with $\hat{H}_0$ refitted under this constraint. 
We fit Eq.~\eqref{eq:fourier_singlemode} to the eight baseline bins using a weighted least-squares procedure (equivalently, maximizing a Gaussian likelihood for the binned $H_0$ measurements), and obtain the best-fit amplitude $A_\star$ and its $1\sigma$ uncertainty $\sigma_A$ from the two-parameter covariance matrix. 
Throughout this work we use
\begin{equation}
N_\sigma \equiv \frac{|A_\star|}{\sigma_A}
\label{eq:Nsigma_def}
\end{equation}
as a single-number measure of the statistical preference for a nonzero smooth oscillatory component in $H_0^{\mathrm{eff}}(z)$ relative to the constant-$H_0$ hypothesis.

\begin{figure}[t]
  \centering
  \includegraphics[width=\columnwidth]{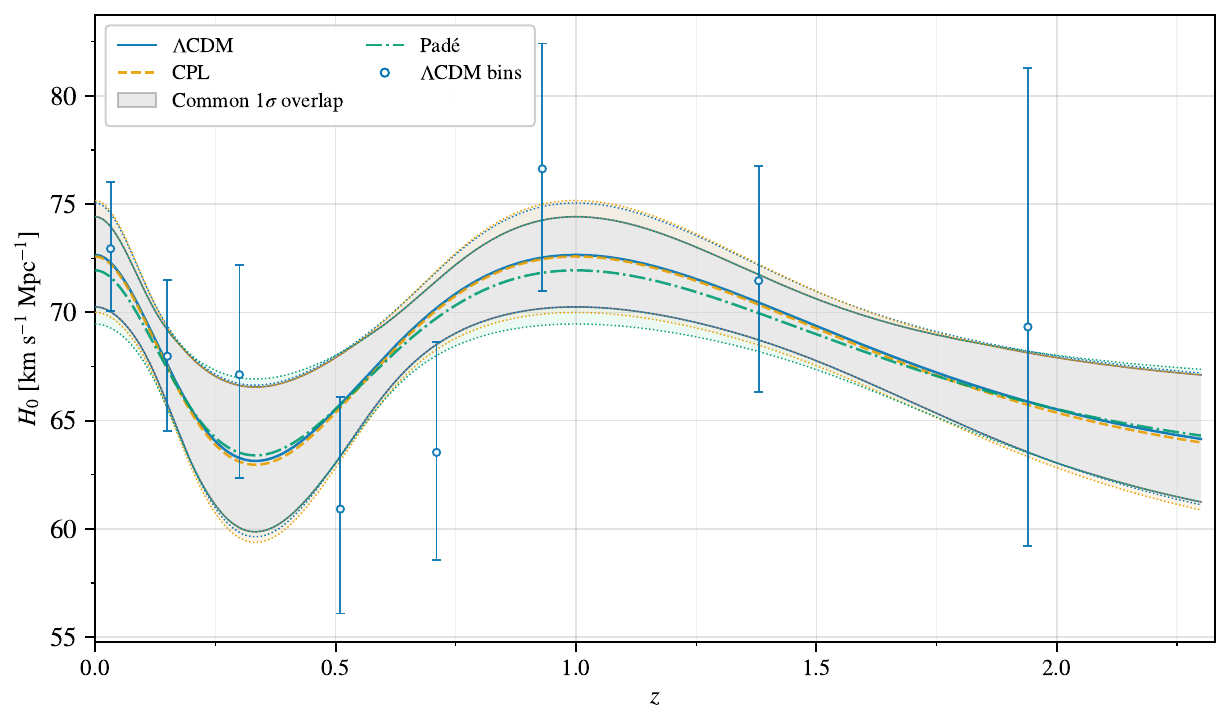}
  \caption{
Fourier-like single-mode fits to the binned $H_0(z)$ obtained in the baseline scheme for the flat $\Lambda$CDM, CPL, and Pad\'e $P_{21}$ models.
Solid, dashed, and dash-dotted curves show the best-fitting
$H_0^{\mathrm{eff}}(z) = \hat{H}_0 + A \cos(4\pi a)$ for $\Lambda$CDM, CPL, and Pad\'e, respectively.
Thin colored dotted lines indicate the individual $1\sigma$ envelopes of each model, while the gray shaded region marks their common $1\sigma$ overlap.
The open circles with error bars denote the baseline $\Lambda$CDM bin measurements, plotted for reference.
  }
  \label{fig:baseline_fourier_fit}
\end{figure}

In practice, we fit Eq.~\eqref{eq:fourier_singlemode} separately to the baseline bins obtained under the flat $\Lambda$CDM, CPL, and Pad\'e models. 
The resulting best-fit parameters are summarized in Table~\ref{tab:fourier_baseline} and shown in Fig.~\ref{fig:baseline_fourier_fit}. 
All three frameworks give mutually consistent values of $\hat{H}_0$ and $A$, with $N_\sigma \simeq 1.71$--$1.94$, indicating that the inferred oscillatory feature is of low statistical significance and largely insensitive to the underlying cosmological parametrization. Compared with Ref.~\cite{LopezHernandez2025}, which studied this behavior in flat CPL cosmologies, our baseline comparison shows that an analogous pattern also arises in flat $\Lambda$CDM and in the kinematical Pad\'e approximation. 

In Sec.~\ref{sec:Analysis} we apply the same diagnostic to alternative binning strategies and to different choices of global versus piecewise-constant parameters, to test the robustness of the oscillatory feature and to assess the role of parameter degeneracies in the apparent evolution of $H_0(z)$.

\begin{table}[b]
\caption{Fourier-like single-mode fits
$H_0^{\mathrm{eff}}(z) = \hat{H}_0 + A\cos(4\pi a)$
to the baseline binned $H_0(z)$. 
Quoted uncertainties are $1\sigma$.
Both $\hat{H}_0$ and $A$ are in units of km\,s$^{-1}$\,Mpc$^{-1}$.}
\label{tab:fourier_baseline}
\begin{ruledtabular}
\begin{tabular}{lccc}
Model         & $\hat{H}_0$        & $A$             & $N_\sigma$ \\
\hline\noalign{\vskip 0.3ex}
Flat $\Lambda$CDM & $67.89 \pm 1.73$ & $4.76 \pm 2.45$ & $1.94$     \\
CPL           & $67.77 \pm 1.81$ & $4.81 \pm 2.54$ & $1.89$     \\
Pad\'e        & $67.66 \pm 1.75$ & $4.27 \pm 2.50$ & $1.71$     \\
\end{tabular}
\end{ruledtabular}
\end{table}

\section[Robustness Tests of Binned H0(z)]{Robustness Tests of Binned $H_0(z)$}
\label{sec:Analysis}
In the previous Section we found that, under the primary binning scheme (A0), the inferred Hubble constant $H_0$ shows a mild redshift-dependent behavior when the eight bins are analysed separately. 
This pattern appears consistently across all three cosmological frameworks.
The goal of this section is to investigate the robustness and possible origin of this effect.

We perform three complementary analyses. 
First, in Sec.~\ref{sec:alt_binning} we repeat the full procedure using three alternative binning schemes (A1--A3), in order to test whether the observed $H_0(z)$ behavior is an artifact of how the data are partitioned in redshift. 
Second, in Sec.~\ref{sec:single_probe} we carry out single-probe fits, in which only one probe (SNe, CC, or BAO) is used at a time to constrain $H_0$ in each bin. 
This allows us to test whether the trend is driven by any single probe, or whether it only emerges when multiple probes are combined. 
Finally, in Sec.~\ref{sec:degen} we consider configurations in which parameters such as $\Omega_m$, the supernova absolute magnitude $M$, and the sound horizon $r_d$ are required to be shared across all bins, while $H_0(z)$ is modeled as a piecewise-constant function of redshift (one constant value per bin). 
This last test directly probes whether the apparent ``evolution'' of $H_0$ can be explained by parameter degeneracies, rather than indicating a genuine physical redshift variation of the Hubble constant.

\subsection{Alternative binning schemes}
\label{sec:alt_binning}

To assess whether the observed redshift-dependent behavior of the Hubble constant $H_0(z)$ depends on the adopted redshift partition, we construct three alternative binning schemes (A1--A3) in addition to the primary configuration A0. 
We use the same datasets and analysis pipeline as in Sec.~\ref{sec:data_method}, changing only the bin boundaries.  

In this Section we restrict to the flat $\Lambda$CDM framework.
As shown in the baseline analysis, the three cosmological frameworks (flat $\Lambda$CDM, CPL, and Pad\'e~(2,1)) yield consistent $H_0(z)$ trends. 
Therefore, we adopt the model with the fewest free parameters and the simplest structure to achieve tighter parameter constraints. 
This choice ensures that the conclusions in this subsection reflect the impact of the binning scheme, rather than any additional sensitivity to model complexity or parameter degeneracies.

\paragraph{Scheme A1: approximately uniform $\Delta z$ binning.}
As a first alternative to the primary scheme A0, we construct a binning in which the full redshift range $z \in [0, 2.33]$ is divided into six bins. 
The first bin is kept fixed as $(0, 0.069]$, identical to the first bin in A0. 
This preserves a dedicated very-low-redshift bin dominated by local distance indicators (nearby SNe~Ia and megamasers), which provides an anchor on the local expansion rate.
The remaining five bins are obtained by partitioning the interval $[0.069, 2.33]$ into five bins of comparable width in redshift. 
After testing different numbers of subdivisions, we find that splitting $[0.069, 2.33]$ into five equal-width bins yields the most stable parameter constraints, while using more bins leads to insufficient statistics per bin.

Unlike in the primary scheme A0, the A1 configuration does not automatically ensure that different observational probes within the same redshift bin share comparable representative redshifts. 
We therefore revise the definition of the effective redshift $z_{\mathrm{eff}}$ for this scheme.

For each dataset $d$ (e.g.\ SNe, CC, or BAO), we first compute its inverse--variance--weighted effective redshift within a given bin,
\begin{equation}
z_{\mathrm{eff},d} = 
\frac{\sum_i z_i / \sigma_i^2}{\sum_i 1 / \sigma_i^2}\,,
\end{equation}
where $z_i$ and $\sigma_i$ denote the redshift and $1\sigma$ uncertainty of each measurement, respectively.  
Then, we combine the effective redshifts of all datasets by weighting them according to their respective number of measurements $N_d$ in that bin:
\begin{equation}
z_{\mathrm{eff,global}} = 
\frac{\sum_d N_d \, z_{\mathrm{eff},d}}{\sum_d N_d}\,,
\qquad
d \in \{\mathrm{SNe},\,\mathrm{CC},\,\mathrm{BAO},\,\mathrm{Maser}\}\, .
\end{equation}

This two-step weighting procedure yields a single representative redshift for each bin that reflects the combined redshift distribution of all probes, even when the effective redshifts of the individual probes are not perfectly aligned.
The same prescription for computing $z_{\mathrm{eff}}$ is also applied to schemes A2 and A3 discussed below.

\paragraph{Scheme A2: CC+BAO quantile binning.}
For A2 we aim to balance the statistical weight of the nonsupernova probes at intermediate and high redshift.
To this end, we construct an eight-bin partition of $z \in [0, 2.33]$.
As in A1, the first bin is fixed to $(0, 0.069]$ in order to preserve a dedicated very-low-$z$ anchor dominated by nearby SNe~Ia and megamasers. 
The remaining seven bins are then defined by the joint distribution of the CC and DESI BAO data points: the bin edges are chosen such that each bin contains, as closely as possible, a comparable total number of CC+BAO measurements.

The motivation for this strategy is that CC and BAO together provide the main late-time expansion-rate and distance information beyond $z \gtrsim 0.07$, independently of the supernova absolute-magnitude calibration. 
However, these probes are much sparser than the supernova sample, and in the primary scheme A0 they tend to cluster around a small number of discrete redshifts (in particular those targeted by BAO). 
This can lead to situations in which one or two bins inherit a disproportionate fraction of the non-SNe information. 
Scheme A2 mitigates this by enforcing a more even CC+BAO sampling over $0.07 \lesssim z \lesssim 2.3$, so that no single redshift interval is dominated by a small set of CC+BAO measurements.

As in scheme A1, each bin in A2 is assigned an effective redshift $z_{\rm eff}$ computed via the same two-step prescription. 
First, we compute an inverse--variance--weighted effective redshift for each dataset within the bin.
Then we take an $N_d$-weighted average over datasets to obtain a single representative $z_{\rm eff}$ for that bin.

\paragraph{Scheme A3: SNe quantile binning.}
For A3 we construct a binning scheme in which the bin edges are chosen to balance the contribution of the Type~Ia supernova sample across redshift.
As in schemes A1 and A2, the first bin is fixed to $(0, 0.069]$ so as to isolate the very low-redshift anchor dominated by nearby SNe~Ia and megamasers. 
The remaining bins are defined by partitioning the interval $z \in [0.069, 2.33]$ into consecutive redshift ranges such that each bin contains approximately the same number of Pantheon+ supernovae.

The motivation for this scheme is twofold. 
First, SNe~Ia are by far the most numerous probe in our compilation and therefore dominate the likelihood at low and intermediate redshift. 
Second, the supernova absolute magnitude $M$ is tightly degenerate with $H_0$, so an uneven redshift distribution of the supernova sample can, in principle, modulate how this degeneracy projects onto the binned $H_0$ estimates.

As in schemes A1 and A2, each bin in A3 is assigned an effective redshift $z_{\rm eff}$ using the same two-step prescription described above.

The redshift ranges, effective redshifts, and data composition for schemes A1--A3 are summarized in Table~\ref{tab:binning_A123}.

\begin{table*}[t]
\centering
\caption{
Redshift ranges, effective redshifts, and data composition for the
alternative binning schemes A1–A3.
Scheme A1 adopts approximately uniform $\Delta z$ bins;
A2 uses quantile binning in CC+BAO;
A3 uses quantile binning in SNe.
Numbers in parentheses indicate the number of data points from each
data set contributing to each bin.
}
\label{tab:binning_A123}
\begin{ruledtabular}
\begin{tabular}{ccccc}
Scheme & Bin & Datasets included & Redshift range & $z_{\rm eff}$ \\
\hline
\multirow{6}{*}{A1}
& 1 & SNe (618), Megamasers (6)
    & $(0.000,\,0.069]$  & 0.032 \\
& 2 & SNe (818), CC (20), DESI--BAO (3)
    & $(0.069,\,0.521]$  & 0.26 \\
& 3 & SNe (162), CC (9), DESI--BAO (4)
    & $(0.521,\,0.973]$  & 0.651 \\
& 4 & SNe (19), CC (4), DESI--BAO (2)
    & $(0.973,\,1.426]$  & 1.211 \\
& 5 & SNe (5), CC (3), DESI--BAO (2)
    & $(1.426,\,1.878]$  & 1.569 \\
& 6 & SNe (2), CC (1), DESI--BAO (2)
    & $(1.878,\,2.330]$  & 2.174 \\
\hline
\multirow{8}{*}{A2}
& 1 & SNe (618), Megamasers (6)
    & $(0.000,\,0.069]$  & 0.032 \\
& 2 & SNe (360), CC (8)
    & $(0.069,\,0.270]$  & 0.180 \\
& 3 & SNe (342), CC (6), DESI--BAO (1)
    & $(0.270,\,0.425]$  & 0.337 \\
& 4 & SNe (102), CC (6)
    & $(0.425,\,0.510]$  & 0.464 \\
& 5 & SNe (167), CC (4), DESI--BAO (4)
    & $(0.510,\,0.781]$  & 0.626 \\
& 6 & SNe (13), CC (5), DESI--BAO (2)
    & $(0.781,\,0.934]$  & 0.841 \\
& 7 & SNe (15), CC (5), DESI--BAO (2)
    & $(0.934,\,1.430]$  & 1.208 \\
& 8 & SNe (7), CC (3), DESI--BAO (4)
    & $(1.430,\,2.330]$  & 1.874 \\
\hline
\multirow{7}{*}{A3}
& 1 & SNe (618), Megamasers (6)
    & $(0.000,\,0.069]$  & 0.032 \\
& 2 & SNe (170), CC (4)
    & $(0.069,\,0.173]$  & 0.125 \\
& 3 & SNe (167), CC (3)
    & $(0.173,\,0.231]$  & 0.199 \\
& 4 & SNe (167), CC (2), DESI--BAO (1)
    & $(0.231,\,0.296]$  & 0.262 \\
& 5 & SNe (167), CC (1)
    & $(0.296,\,0.371]$  & 0.332 \\
& 6 & SNe (167), CC (10), DESI--BAO (2)
    & $(0.371,\,0.549]$  & 0.447 \\
& 7 & SNe (168), CC (17), DESI--BAO (8)
    & $(0.549,\,2.261]$  & 0.754 \\
\end{tabular}
\end{ruledtabular}
\end{table*}

\paragraph{Fourier fit to alternative binning schemes}
To test how sensitive the apparent $H_0(z)$ evolution is to the choice of redshift binning, we apply the Fourier-like diagnostic introduced in Sec.~\ref{sec:fourier} to the four binning schemes A0--A3 (within the flat $\Lambda$CDM framework). 
For each scheme, we fit the single-mode ansatz of Eq.~\eqref{eq:fourier_singlemode} to the corresponding set of binned $H_0$ constraints, using a weighted least-squares fit.
As in Sec.~\ref{sec:fourier}, the fitted $(\hat{H}_0, A)$ provide a compact summary of the effective trend and its amplitude under each binning prescription.

\begin{figure*}[t]
    \centering
    \includegraphics[width=1.0\textwidth]{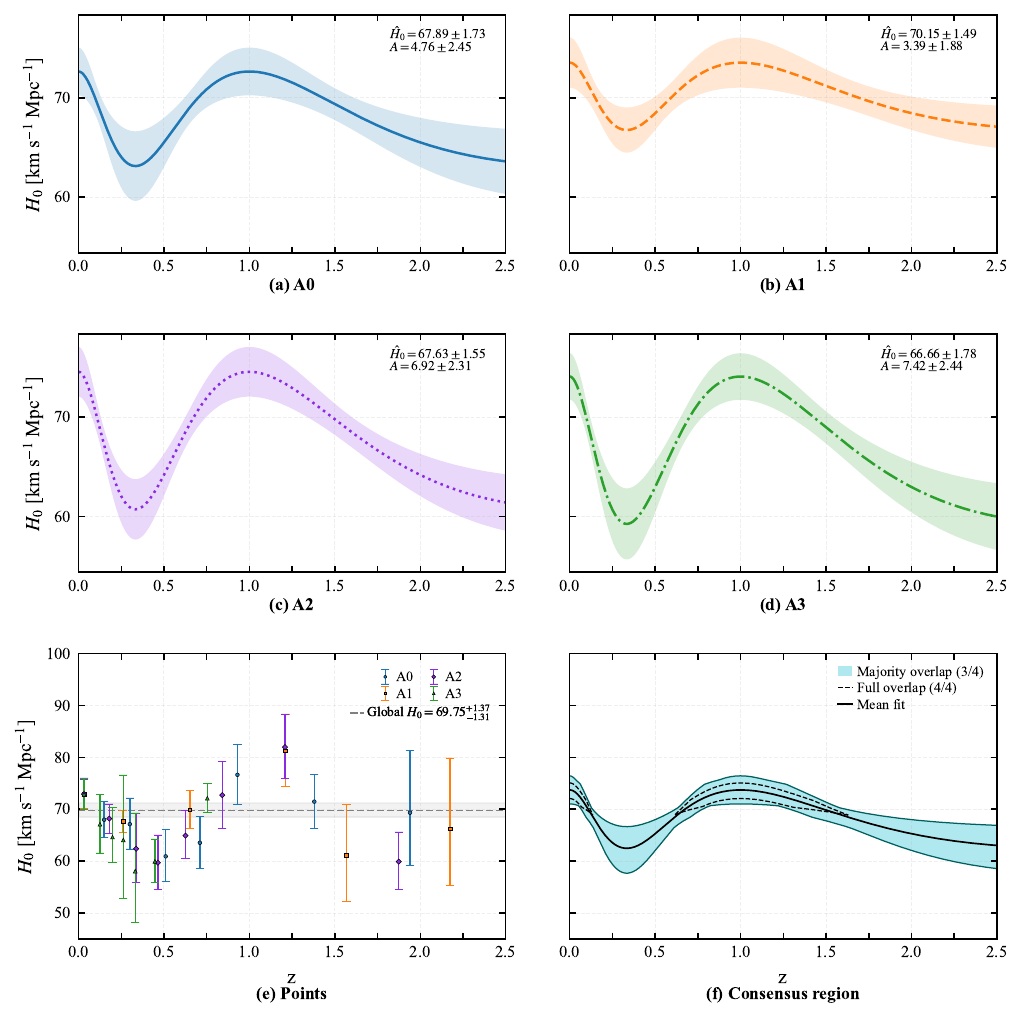}
    \caption{
    Alternative binning scheme results. 
    (a--d) For each binning scheme (A0–A3), the solid curve shows the Fourier-like best-fit
    trend of $H_0(z)$ and the shaded band indicates its $1\sigma$ credible region; the upper-left annotations list
    the fitted $(\hat H_0, A)$ values. 
    (e) Points-only comparison: each marker is the binned $H_0$ at the effective
    redshift $z_{\rm eff}$ with a $1\sigma$ error bar; the gray band indicates the global (unbinned) estimate from the
    same data and model. 
    (f) Consensus region across schemes: the shaded area denotes the majority
    overlap (3/4), the black dashed lines mark the full overlap (4/4), and the solid black curve is the
    mean of the four fitted trends.
    }
    \label{fig:results4p1}
\end{figure*}

Figure~\ref{fig:results4p1} shows the Fourier-like fits and their $1\sigma$ credible regions for each binning scheme (panels~a--d), the points-only comparison (panel~e), and a consensus view across schemes (panel~f). For interpretation in Fig.~\ref{fig:results4p1}, we define the majority overlap (3/4) as the redshift intervals where at least three of the four schemes’ $1\sigma$ bands overlap, and the full overlap (4/4) as the intersection common to all four $1\sigma$ bands. The results indicate a consistent oscillatory pattern in $H_0(z)$ across schemes: $H_0(z)$ decreases over $z \simeq 0$--$0.45$ to a broad minimum, rises again over $z \simeq 0.45$–$1.0$ to a local maximum, and then gently declines toward a nearly stable high-$z$ plateau. The overall oscillatory feature amplitude is at the level of a few ${\rm km\,s^{-1}\,Mpc^{-1}}$. 

In contrast, scheme A1 (equal $\Delta z$) partitions the data into fewer, unevenly populated bins and, unlike the other binning strategies, does not place any effective redshift close to the location of the global minimum of the trend. As a result, the Fourier-like fit in A1 is less sensitive to the trough of the oscillation, yielding a weaker and less pronounced amplitude. Scheme A3 (equal-count SNe), whose effective redshifts are concentrated in the low-to-intermediate region ($0.03 \lesssim z \lesssim 0.75$), produces a deeper minimum and hence a slightly larger amplitude. Quantitatively, the amplitudes inferred from A0–A3 are mutually consistent within $\lesssim 1.3\sigma$, so none of the schemes stands out as a statistically significant outlier.

Overall, despite the differences in bin boundaries and redshift sampling, the inferred $H_0(z)$ trends remain qualitatively similar across all schemes. This indicates that the apparent evolution of the Hubble constant with redshift is not an artifact of a specific binning choice, but a stable, data-driven feature.

\subsection{Single-probe binned fits}
\label{sec:single_probe}
In the baseline scheme introduced in Sec.~\ref{sec:results}, each bin is constrained by a joint likelihood that combines at least two distinct observational probes. As a consequence, the apparent behavior of $H_0(z)$ in that scheme may not correspond to a feature preferred by any single probe in isolation; it could instead arise as a coupled, multiprobe effect driven by the interplay between probes and their parameter degeneracies. To assess this possibility, we repeat the binned inference for each probe separately: we keep the same redshift binning (A0) and the same flat $\Lambda$CDM model (for the reasons discussed in Sec.~\ref{sec:alt_binning}), but now fit only one probe at a time (SNe only, CC only, and BAO only).

Our aim is to test whether the $H_0(z)$ variation seen in the baseline scheme persists within any single probe when analyzed in isolation. If a similar oscillatory pattern is recovered independently from the individual probes, this would indicate that the apparent behavior of $H_0(z)$ is a robust feature supported internally by the data.

\paragraph{Pantheon+ (SNe only).}
In this work we do not apply any external distance-ladder calibration to fix the absolute magnitude $M$ of Type~Ia supernovae. As a consequence, Pantheon+ alone cannot determine $H_0$ directly; instead it constrains only the degeneracy between $H_0$ and $M$. This can be seen by rewriting the theoretical model for the supernova apparent magnitude.

In the baseline scheme, the predicted apparent magnitude of a supernova at redshift $z$ can be written as
\begin{equation}
m_B^{\rm th}(z)
= M + 5 \log_{10} \left[ \frac{c (1+z)}{H_0}
\int_0^{z} \frac{dz'}{E(z';\Omega_m)} \right] + 25,
\label{eq:SN_mag_base}
\end{equation}
where $E(z) \equiv H(z)/H_0$ is the dimensionless Hubble rate. Factoring $H_0$ out of the luminosity-distance factor gives
\begin{equation}
m_B^{\rm th}(z)
= M' + 5 \log_{10} \left[ c (1+z)
\int_0^{z} \frac{dz'}{E(z';\Omega_m)} \right] + 25,
\label{eq:SN_mag_reparam}
\end{equation}
where we have defined
\begin{equation}
M' \equiv M - 5 \log_{10} H_0.
\label{Mpri}
\end{equation}
In this reparameterization, $H_0$ does not appear explicitly: the fit is performed in terms of $M'$, which encodes the $H_0$--$M$ degeneracy.

Following previous supernova-only binning analyses (e.g.\ Refs.~\cite{Dainotti2021,Montani2025}), we adopt a two-step procedure.
First, we run an MCMC analysis on the full Pantheon+ sample (without binning) under flat $\Lambda$CDM and obtain a global constraint on the matter density, which we denote $\Omega_{m,{\rm global}}$. Second, when fitting each individual redshift bin, we fix $\Omega_m = \Omega_{m,{\rm global}}$ and fit only for $M'$. Thus each bin is effectively a one-parameter inference in which $M'$ captures how the $H_0$--$M$ combination preferred by the supernovae changes with redshift.

This construction implies that Pantheon+ alone cannot yield an absolute value of $H_0$ in each bin. However, any systematic drift of $M'$ across the bins serves as the supernova-only analog of an $H_0(z)$ variation.

\paragraph{Cosmic chronometers (CC only).}
Cosmic chronometer measurements provide a direct estimate of the expansion rate $H(z)$ from galaxy ages. As discussed in Sec.~\ref{sec:data}, we use 37 CC measurements, which do not include any points in the first redshift bin of the A0 scheme; hence only bins 2--8 can be constrained.

We apply the same two-step procedure as for the supernova sample. First, we perform an MCMC analysis of the full CC catalog under flat $\Lambda$CDM to obtain a global matter-density estimate. Second, when fitting each individual redshift bin (bins 2--8) we fix $\Omega_m = \Omega_{m,{\rm global}}$ and fit only for $H_0$ using the CC likelihood defined in Sec.~\ref{sec:likelihood}. Thus, for CC we obtain a direct per-bin constraint on $H_0$ itself.

\paragraph{BAO only.}
For BAO we consider the same DESI~DR2 measurements described in Sec.~\ref{sec:data}, including both isotropic ($D_V/r_d$) and anisotropic ($D_M/r_d$, $D_H/r_d$) observables. In flat $\Lambda$CDM the corresponding theoretical quantities are
\begin{align}
\frac{D_H(z)}{r_d} &= \frac{c}{H(z)\, r_d}
= \frac{c}{(H_0 \, r_d)} \, \frac{1}{E(z;\Omega_m)}, \\
\frac{D_M(z)}{r_d} &= \frac{c}{(H_0 \, r_d)}
\int_0^{z} \frac{dz'}{E(z';\Omega_m)}, \\
\frac{D_V(z)}{r_d} &= 
\left[
z \left( \frac{D_M(z)}{r_d} \right)^2
\frac{D_H(z)}{r_d}
\right]^{1/3} .
\end{align}
All of these quantities depend on $\Omega_m$ and on the product $H_0 r_d$, rather than on $H_0$ and $r_d$ separately. Consequently, BAO alone cannot break the degeneracy between $H_0$ and $r_d$.

In the A0 binning scheme the BAO sample has no coverage in the first two redshift bins, so BAO-only constraints can be derived only for bins 3--8. We analyze BAO in close analogy to the CC and SNe cases. First, we perform an MCMC analysis of the full DESI BAO sample (without binning) under flat $\Lambda$CDM to obtain a global matter-density estimate. Second, for each individual BAO bin (bins 3--8) we fix $\Omega_m = \Omega_{m,{\rm global}}$ and fit only a single free parameter describing that bin, namely the combination
\begin{equation}
(H_0 r_d)_{\rm bin} \equiv H_0 \, r_d .
\end{equation}
Thus, for BAO we do not recover an absolute $H_0$ per bin; instead we obtain the variation of $H_0 r_d$, which plays the same diagnostic role for BAO as $M'$ does for the supernova sample.

\paragraph{Comparison.}
To compare single--probe fits (SNe only, CC only, BAO only) with the baseline scheme on a common, dimensionless scale, we express each constraint as a ratio to the global (unbinned) estimate of the same probe. Concretely, we use \(H_{0,i}^{\rm (CC)}/H_{0,\mathrm{global}}^{\rm (CC)}\) and \(H_{0,i}^{\rm (base)}/H_{0,\mathrm{global}}^{\rm (base)}\); for SNe, Eq.~\eqref{Mpri} gives \(H_{0,i}^{\rm (SNe)}/H_{0,\mathrm{global}}^{\rm (SNe)} = 10^{(M'_{\mathrm{global}}-M'_i)/5}\); and for BAO,
\(H_{0,i}^{\rm (BAO)}/H_{0,\mathrm{global}}^{\rm (BAO)} = (H_0 r_d)_i^{\rm (BAO)}/(H_0 r_d)_{\mathrm{global}}^{\rm (BAO)}\). 
This construction is based on a simple physical assumption: within each probe, the calibration parameters --- \(M\) for SNe and \(r_d\) for BAO --- do not vary across redshift bins and are consistent with their global estimates. We do not fix numerical values for \(M\) or \(r_d\), and neither do we focus on absolute \(H_0\) for each probe. Instead, we use dimensionless ratios that remove these calibrations and isolate the relative redshift variation of \(H_0(z)\). This enables a calibration--independent cross--probe comparison.

\begin{figure}[t]
  \centering
  \includegraphics[width=\columnwidth]{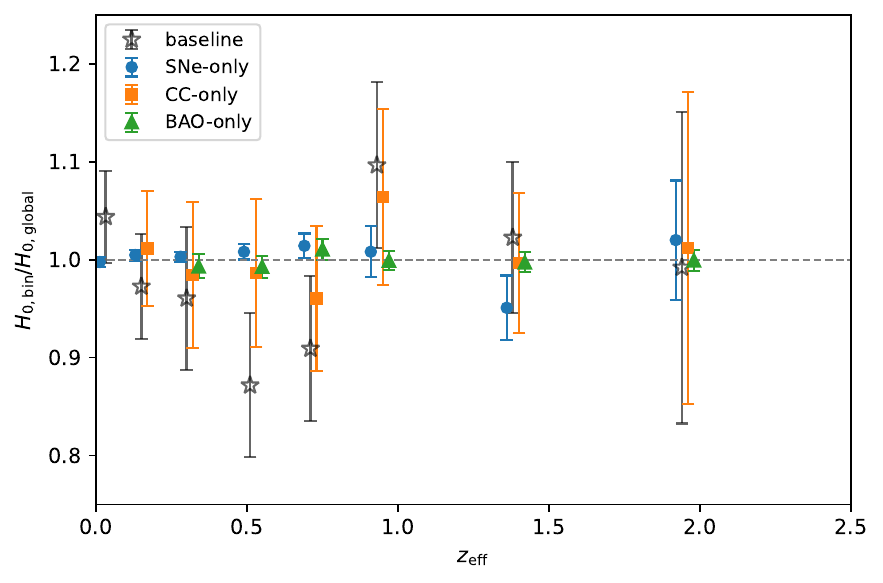}
  \caption{
Per-bin ratios of the Hubble constant, $H_{0,i}/H_{0,\mathrm{global}}$, versus effective redshift $z_{\rm eff}$, for the single-probe binned fits compared with the baseline scheme.
Hollow five-point stars denote the baseline scheme; filled circles, squares, and triangles correspond to SNe only, CC only, and BAO only, respectively.
Vertical bars show $1\sigma$ uncertainties.
The dashed horizontal line marks $H_{0,i}/H_{0,\mathrm{global}} = 1$, i.e.\ no deviation from each probe's global scale.
  }
  \label{fig:4p2results}
\end{figure}

Figure~\ref{fig:4p2results} displays the per-bin ratios \(H_{0,i}/H_{0,\mathrm{global}}\) for the baseline (hollow stars) and for the three single-probe fits (SNe only, CC only, BAO only). For each probe, points lie within their \(1\sigma\) uncertainties of unity, indicating no compelling monotonic variation once the constraints are expressed as ratios. Comparing the single-probe series with the baseline, none of the individual probes reproduces the pronounced baseline pattern seen in the joint fit. This suggests that the apparent behavior in the baseline is not driven by any single probe, but more plausibly reflects the coupling of probe--specific parameter degeneracies (e.g., \(H_0\)–\(M'\) for SNe and \(H_0\)–\(r_d\) for BAO) within the combined likelihood. 

Moreover, the three single-probe curves do not exhibit a coherent common pattern: their ratios fluctuate around unity with no shared direction across bins. If a genuine, probe-independent redshift evolution of \(H_0\) were present, one would expect aligned departures from unity across all three probes. The absence of such coherence therefore does not support a strong universal evolution within current uncertainties and binning.

\subsection{Degeneracy analysis}
\label{sec:degen}

In the baseline binned results (Sec.~\ref{sec:results}), we find that the inferred $H_0(z)$ exhibits an oscillatory behavior whose significance reaches a marginal level of $\sim 2\sigma$ when quantified using the Fourier parametrization. We also observe that other parameters constrained in our MCMC analysis, such as the supernova absolute magnitude $M$ and the sound horizon $r_d$, display indications of oscillatory behavior across the bins at a comparable level.
These findings motivate us to test whether the apparent redshift-dependent variations can be explained by underlying parameter degeneracies.

To clarify the statistical origin of these patterns, we first quantify the strength and orientation of parameter degeneracies using the sample covariance of the MCMC chains computed per bin from the post--burn--in samples.
The degeneracy strength between parameters $x$ and $y$ is characterized by the Pearson correlation coefficient
\begin{equation}
\rho_{x,y} \;=\; \frac{\mathrm{Cov}(x,y)}{\sigma_x\,\sigma_y},
\label{eq:rho}
\end{equation}
where $\mathrm{Cov}(x,y)$ is the sample covariance in that bin, and $\sigma_x$ and $\sigma_y$ are the corresponding
sample standard deviations.
The orientation of the joint posterior ellipse in the $(x,y)$ plane is described by
\begin{equation}
\tan\!\bigl(2\theta\bigr) \;=\; \frac{2\,\mathrm{Cov}(x,y)}{\sigma_x^{2}-\sigma_y^{2}},
\label{eq:theta}
\end{equation}
which follows from the eigen-decomposition of the $2{\times}2$ covariance matrix.
Equations~\eqref{eq:rho} and \eqref{eq:theta} together quantify both how strongly two parameters are coupled and how the direction of that coupling varies across redshift bins.

Using the baseline flat $\Lambda$CDM configuration, we illustrate these relationships in Appendix~\ref{app:degen_figs}, where the joint posteriors of $(H_0,\Omega_m)$, $(H_0,M)$, and $(H_0,r_d)$ from the eight redshift bins are overplotted, along with tables summarizing the corresponding $\rho$ and $\theta$ values. 
These visualizations show that $H_0$--$M$ and $H_0$--$r_d$ are strongly correlated in the SNe- and BAO-dominated bins, respectively, while the $H_0$--$\Omega_m$ degeneracy direction rotates systematically with redshift. 
Such behavior suggests that the apparent oscillatory behavior of $H_0(z)$ may not reflect a genuine redshift evolution of the expansion rate, but rather is closely linked to parameter degeneracies among $\{H_0,\Omega_m,M,r_d\}$.

To test this hypothesis quantitatively, we extend the baseline binned analysis by constructing a family of piecewise-constant models with the same binning, datasets, and priors. In contrast to the independent per-bin fits of Sec.~\ref{sec:results}, these extensions are realized through a single global MCMC analysis, in which the Hubble constant is modeled as a redshift-dependent, piecewise-constant function $H_0(z)$,
\begin{equation}
H(z) =
\begin{cases}
H_{0,1}\,E(z;\Omega_m), & z \in \text{bin 1}, \\
H_{0,2}\,E(z;\Omega_m), & z \in \text{bin 2}, \\
\vdots \\
H_{0,K}\,E(z;\Omega_m), & z \in \text{bin K},
\end{cases}
\label{eq:piecewise_H}
\end{equation}
where $E(z;\Omega_m)$ is the normalized expansion function and $K$ denotes the
number of bins.
In the simplest configuration, all other cosmological and calibration parameters ($\Omega_m$, $M$, $r_d$) are treated as global and shared across all bins.

The total likelihood is defined as the product of the likelihood contributions
from all probes and all bins (or equivalently, the total $\chi^2$ is the sum of
their contributions):
\begin{align}
\ln \mathcal{L}(\boldsymbol{\theta})
= -\frac{1}{2}\Big[
 & \chi^2_{\mathrm{SNe}}(H_{0,k},\Omega_m,M)
 + \chi^2_{\mathrm{BAO}}(H_{0,k},\Omega_m,r_d)
 \nonumber\\
 & + \chi^2_{\mathrm{CC}}(H_{0,k},\Omega_m)
 + \chi^2_{\mathrm{maser}}(H_{0,1},\Omega_m)
\Big],
\label{eq:global_like}
\end{align}
where $\boldsymbol{\theta}$ denotes the full parameter vector, and each term is
evaluated using the appropriate $H_{0,k}$ according to the bin assignment of
each data point.
In this setup, the piecewise-constant $H_0(z)$ enters directly in the joint likelihood, so that all segments are constrained simultaneously in a single global fit.

We stress that treating a parameter as global and shared across bins is conceptually distinct from fixing its value independently in each bin. In our framework, parameters other than the Hubble constant---such as \(\Omega_m\), \(M\), and \(r_d\)---are left free in the global fit and are simply required to take the same value across all redshift intervals. By contrast, only \(H_0\) is allowed to vary in a piecewise-constant manner. This setup helps us test whether the observed oscillatory behaviour of \(H_0(z)\) reflects a genuine redshift evolution of the expansion rate. If the pattern weakens or disappears once \(\Omega_m\), \(M\), and \(r_d\) are enforced to be global, this would indicate that the baseline behavior is primarily produced by degeneracy freedom of these parameters, i.e. by shifts along their joint degeneracy directions with \(H_0\).

Having established the piecewise-constant $H_0(z)$ framework, we next extend this idea by progressively allowing additional parameters that are known to be degenerate with $H_0$ --- namely $\Omega_m$, $M$, and $r_d$ --- to vary as independent constants in each redshift bin.
In each extended configuration, the selected parameter(s) adopt the same piecewise-constant structure as $H_0(z)$ and enter the likelihood accordingly.
This leads to a sequence of binned-parameter models in which only $H_0$, or parameter sets such as $\{H_0,\Omega_m\}$, $\{H_0,M\}$, $\{H_0,r_d\}$, and finally
all four parameters $\{H_0,\Omega_m,M,r_d\}$, are promoted from global to bin-dependent, piecewise constant quantities.
All these models share the same global likelihood structure as Eq.~\eqref{eq:global_like}; they differ only in which parameters are treated as piecewise-constant across bins.

\begin{figure*}[t]
  \centering
  \includegraphics[width=\textwidth]{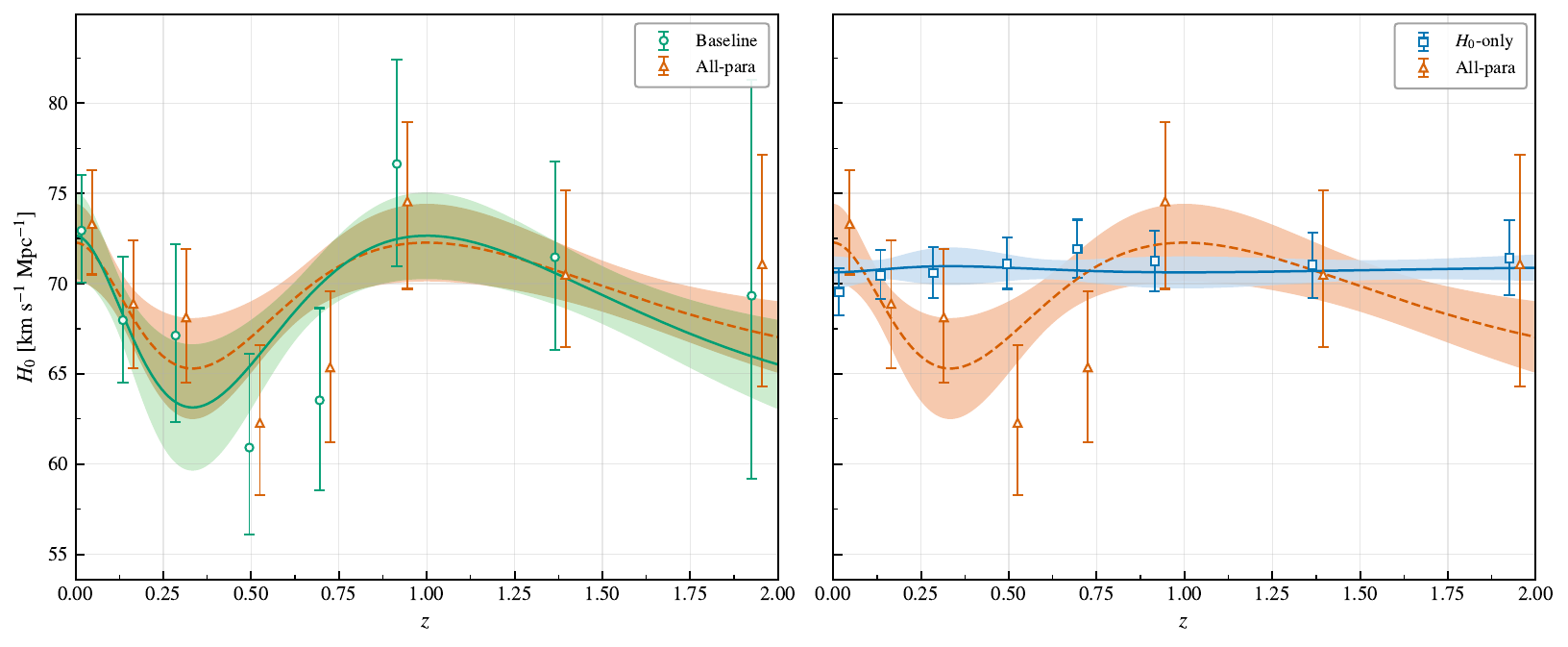}
  \caption{
Constraints on piecewise-constant $H_0(z)$ for different parameter configurations.
The left panel compares the baseline binned analysis (baseline) with the case in which all four parameters $\{H_0,\Omega_m,M,r_d\}$ are treated as piecewise constant (All-para).
The right panel shows the configuration in which only $H_0$ varies between bins ($H_0$ only), compared with the All-para case.
In both panels, the shaded bands show the $1\sigma$ uncertainty of the corresponding Fourier fits, with colors matching the associated data points.
  }
  \label{fig:4p3results}
\end{figure*}

For each piecewise-constant configuration, we use the global MCMC constraints on the segment parameters $\{H_{0,k}\}_{k=1}^K$ and fit them with the single-mode Fourier form of Eq.~\eqref{eq:fourier_singlemode}. This allows us to compare their redshift-dependent behavior and to quantify the corresponding oscillation amplitudes.
The best-fit parameters of the Fourier model for the baseline and all piecewise-constant configurations are summarized in Table~\ref{tab:4p3results}.
In Fig.~\ref{fig:4p3results} we compare the reconstructed $H_0(z)$ from the baseline analysis with that from the fully binned configuration in
which all $\{H_0,\Omega_m,M,r_d\}$ are allowed to vary independently across bins, together with their corresponding Fourier fits.

\begin{table}[t]
\centering
\caption{
Best-fit parameters of the Fourier model for the effective $H_0(z)$
obtained from the baseline analysis and the piecewise-constant
configurations discussed in Sec.~\ref{sec:degen}.
“Baseline’’ denotes the independent binned fit; “$H_0$ only’’ allows
only $H_0$ to vary between bins while all other parameters are kept
global; “$H_0,\Omega_m$’’, “$H_0,M$’’, and “$H_0,r_d$’’ denote
configurations in which $H_0$ and, respectively, $\Omega_m$, $M$, or
$r_d$ are treated as piecewise-constant in redshift; “All-para’’
denotes the configuration in which all four parameters
$\{H_0,\Omega_m,M,r_d\}$ are piecewise constant.
Both $\hat{H}_0$ and $A$ are in units of km\,s$^{-1}$\,Mpc$^{-1}$.}
\label{tab:4p3results}
\begin{ruledtabular}
\begin{tabular}{lcc}
Model         & $\hat{H}_0$        & $A$             \\
\hline
Baseline       & $67.59 \pm 1.76$  & $4.76 \pm 2.45$ \\
$H_0$ only     & $70.79 \pm 0.55$  & $-0.17 \pm 0.79$ \\
$H_0,\Omega_m$ & $70.35 \pm 0.77$  & $-0.02 \pm 1.05$ \\
$H_0,M$        & $69.86 \pm 0.82$  & $0.89 \pm 1.17$ \\
$H_0,r_d$      & $70.49 \pm 0.57$  & $-0.37 \pm 0.83$ \\
All-para       & $68.78 \pm 1.45$  & $3.49 \pm 2.30$ \\
\end{tabular}
\end{ruledtabular}
\end{table}

We find that the configuration in which all four parameters $\{H_0,\Omega_m,M,r_d\}$ are treated as piecewise-constant in redshift yields an effective $H_0(z)$ curve that closely matches the baseline result, both in its overall oscillatory pattern and in the redshift bins where the highest and lowest binned $H_0$ values occur. This shows that the baseline-like oscillatory behavior can be reproduced within a fully piecewise-constant, globally fitted framework, demonstrating the internal consistency of our piecewise-constant construction. In this fully flexible setup, the Fourier amplitude of the oscillatory pattern is moderately reduced compared with the baseline analysis (from $A \simeq 4.76$ to $A \simeq 3.49$; see Table~\ref{tab:4p3results}). This reduction is expected, because all piecewise-constant parameter values are constrained simultaneously by the global likelihood, which induces correlations between bins and penalizes large bin-to-bin deviations from the overall trend.

By contrast, when only $H_0$ is treated as piecewise-constant while $\Omega_m$, $M$, and $r_d$ are kept global, the oscillation is strongly suppressed and the resulting $H_0(z)$ is consistent with a nearly constant value. Combining this with the restricted configurations in which $H_0$ plus a single additional parameter are treated as piecewise-constant, we find that all these partial piecewise constant models yield Fourier amplitudes that remain below the $1\sigma$ level. This indicates that a baselinelike oscillatory behavior reemerges only when the full set $\{H_0,\Omega_m,M,r_d\}$ is allowed to vary between bins. 

When either $M$ or $r_d$ is also treated as bin dependent, the Fourier amplitudes become slightly larger than in the $H_0,\Omega_m$ configuration. This is consistent with the expectation that the $H_0$--$M$ and $H_0$--$r_d$ degeneracies are more strongly correlated with $H_0$. However, given the current uncertainties on the amplitudes, these results should be regarded as suggestive rather than as a robust quantitative ranking.

Overall, this comparison is consistent with interpreting the oscillatory feature as being driven by the available parameter freedom along existing degeneracy directions, rather than as robust evidence for a genuine redshift evolution of the expansion rate.

\section{Conclusions and discussion}
\label{sec:conclusions}

In light of recent indications that the Hubble constant may exhibit redshift-dependent behavior, we have re-examined whether the apparent oscillatory pattern seen in binned late-time measurements of \(H_0(z)\) reflects a genuine feature of the expansion history or instead arises from cosmological-model choices and statistical methodology. To this end, we employ a unified framework that varies the cosmological parametrization, the binning strategy, the probe combination, and the parameter-sharing assumptions.

In the baseline scheme (Sec.~\ref{sec:results}), and for all three cosmological models considered (flat \(\Lambda\)CDM, CPL and Pad\'e), the binned \(H_0\) constraints display a highly consistent oscillatory behavior with redshift. A simple Fourier-like fit to \(H_0(z)\) yields a nonzero oscillatory amplitude with a marginal significance of about \(1.71\text{--}1.94\sigma\), depending on the cosmological model.
To assess the robustness and possible origin of this behavior, we have examined three aspects within the same framework: alternative redshift binning, single-probe per-bin fits, and different global versus piecewise-constant configurations for \(\{H_0,\Omega_m,M,r_d\}\).

We therefore reach the following practical conclusions. First, the oscillatory behaviour of \(H_0(z)\) is present in current late-time data at only marginal (\(\sim 2\sigma\)) significance, but is broadly stable against changes in both the cosmological model and the redshift-binning strategy. Second, the absence of a coherent, probe-independent departure in the single--probe per-bin results and the strong coupling between \(H_0\) and calibration parameters such as \(M\) and \(r_d\) imply that the baseline-level oscillatory trend is plausibly explained by probe-specific degeneracies. When these degeneracies are combined in a joint likelihood and given sufficient parameter freedom, they produce an apparent redshift-dependent pattern in the inferred \(H_0\) constraints. Once the degeneracy freedom between \(H_0\) and other key parameters is restricted (as in the configuration where only \(H_0\) is allowed to vary between bins), the oscillatory trend is strongly suppressed. The inferred values then become consistent with an approximately constant \(H_0 \simeq 70.79\,\mathrm{km\,s^{-1}\,Mpc^{-1}}\) across redshift.

In this context, it is useful to place our findings in relation to previous studies that reported similar features. In particular, Ref.~\cite{LopezHernandez2025} identified an oscillatory pattern in binned late-time measurements of \(H_0\) and argued that this behaviour provides further evidence that the Hubble tension is not restricted to an early--versus--late-time inconsistency, but also manifests as a mild, internal discrepancy among purely late-time determinations. We confirm that a Fourier-like oscillatory trend at the \(\sim 2\sigma\) level can indeed emerge from current late-time data. However, our multimodel and robustness analysis indicates that this pattern is more naturally interpreted as a manifestation of parameter freedom along existing degeneracy directions, rather than as robust evidence for new physics or genuine redshift evolution of the expansion rate.
This interpretation does not dismiss the global Hubble tension, but suggests that the mild internal features highlighted in previous binned \(H_0(z)\) analyses should be treated with caution until parameter degeneracies and late-time systematics are more tightly controlled.

It is worth noting that the analysis assumes that key calibration and cosmological parameters such as the supernova absolute magnitude \(M\), the sound horizon \(r_d\), and \(\Omega_m\) do not exhibit unmodelled redshift evolution across the probed range. Any residual redshift-dependent systematics or genuine evolution in these quantities would, within our framework, be absorbed into the per-bin \(H_0\) estimates, so that their contribution cannot be cleanly disentangled from the apparent oscillatory trend.
In addition, the statistical power is limited by the current data volume and precision, especially in the highest-redshift bins where data are sparse and uncertainties are large. The detailed shape of \(H_0(z)\) at high \(z\) should therefore not be overinterpreted.
A possible next step is to use dedicated sensitivity studies and realistic mock-data analyses to quantify how the main degenerate parameter combinations of each probe combine in joint constraints to shape the inferred \(H_0(z)\).

Numerical analyses were carried out in \texttt{PYTHON}, making use of the \texttt{EMCEE} MCMC sampler and other open-source scientific packages.

\begin{acknowledgments}
We thank Yu-Chen Wang, Bo Yu, Jing Niu, and Wei Hong for valuable discussions and helpful suggestions. 
This work was supported by the National SKA Program of China (No.~2022SKA0110202); 
the China Manned Space Program (CMS-CSST-2025-A01); 
the National Natural Science Foundation of China (NSFC) Youth Program (No.~12403004); 
the China Postdoctoral Science Foundation Postdoctoral Fellowship Program (Grade~C, No.~GZC20241563); 
and the National Key Program for Science and Technology Research and Development (No.~2023YFB3002500).
\end{acknowledgments}

\section*{DATA AVAILABILITY}
The data that support the findings of this article are openly available~\cite{Mo2026Zenodo}.

\appendix
\section{SUPPLEMENTARY DEGENERACY PLOTS AND TABLES}
\label{app:degen_figs}
\begin{figure*}[!t]
\centering
\includegraphics[width=0.32\textwidth]{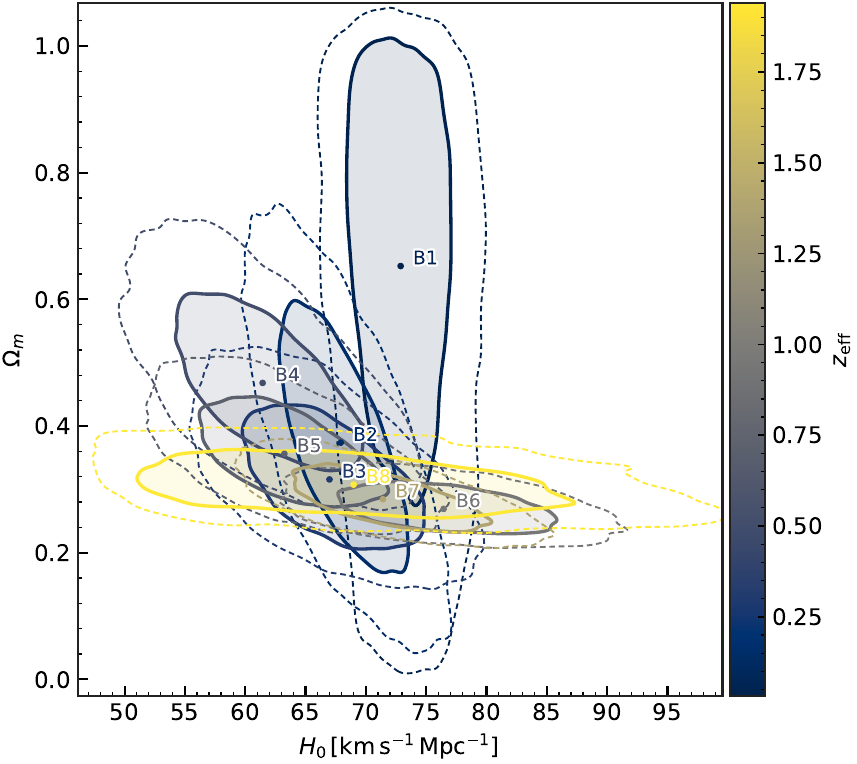}
\includegraphics[width=0.32\textwidth]{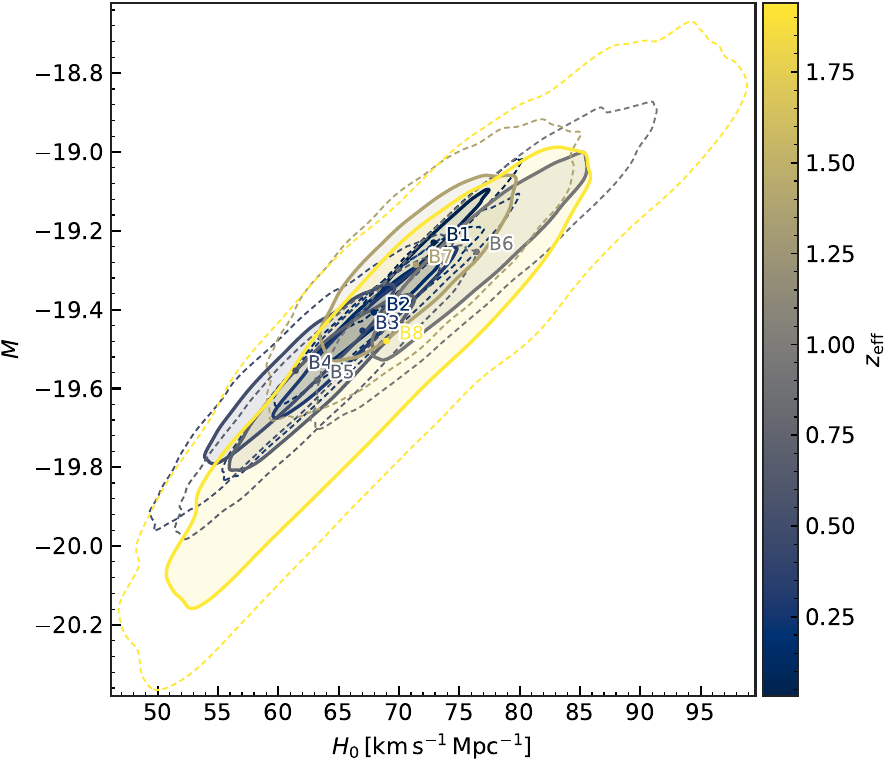}
\includegraphics[width=0.32\textwidth]{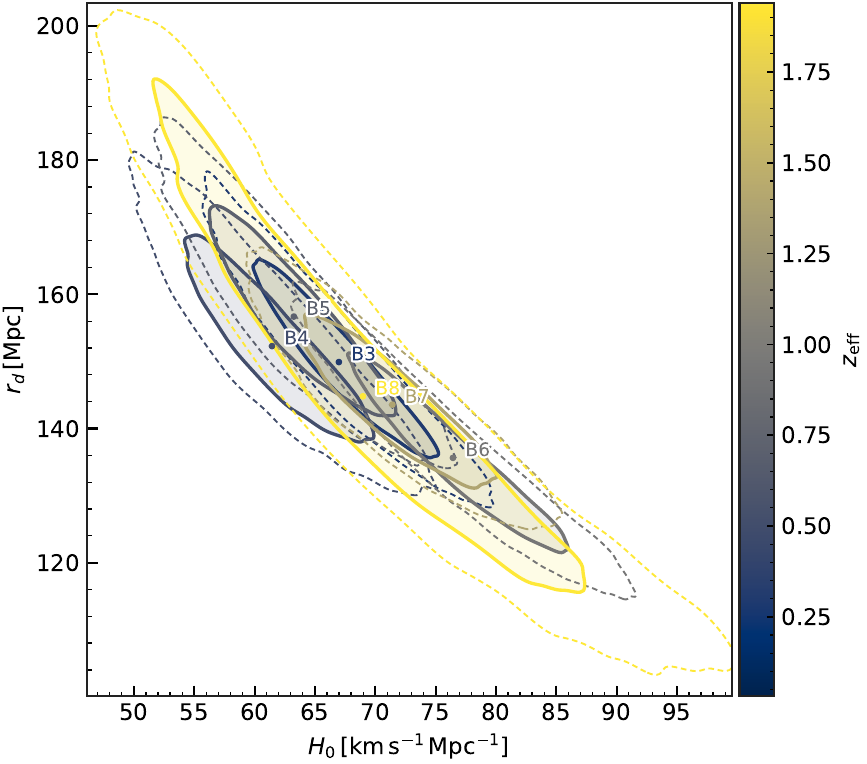}
\caption{
Joint posterior distributions illustrating the parameter degeneracies between
$H_0$ and $(\Omega_m, M, r_d)$ across the eight redshift bins under the
baseline flat $\Lambda$CDM configuration.
Contours show 68\% and 95\% credible regions, color-coded by the effective redshift $z_{\rm eff}$.
The $H_0$--$M$ and $H_0$--$r_d$ correlations are strong in the SNe- and BAO-dominated bins, respectively, while the $H_0$--$\Omega_m$ degeneracy direction rotates systematically with redshift.
}
\label{fig:appendix_degen}
\end{figure*}

To complement Sec.~\ref{sec:degen}, this Appendix shows the two-parameter posterior contours that display the degeneracy structure among $\{H_0,\Omega_m,M,r_d\}$ across the eight redshift bins under the baseline flat
$\Lambda$CDM configuration (Fig.~\ref{fig:appendix_degen}), and provides a compact summary of the corresponding Pearson coefficients $\rho$ and ellipse orientation angles $\theta$ (Table~\ref{tab:degen_summary}).

\begin{table*}[!t]
\caption{
Pearson correlation coefficients ($\rho$) and ellipse orientation angles ($\theta$)
for $(H_0,\Omega_m)$, $(H_0,M)$, and $(H_0,r_d)$ in each redshift bin under the
baseline flat $\Lambda$CDM configuration. Angles are in degrees.
Missing entries in $(H_0,r_d)$ for bins~1--2 reflect the lack of BAO leverage at very low redshift.
}
\label{tab:degen_summary}
\begin{ruledtabular}
\begin{tabular}{cccccccc}
\multirow{2}{*}{Bin} & \multirow{2}{*}{$z_{\rm eff}$} &
\multicolumn{3}{c}{$\rho$} &
\multicolumn{3}{c}{$\theta$ [deg]} \\
\cline{3-5}\cline{6-8}
 & & $H_0$--$\Omega_m$ & $H_0$--$M$ & $H_0$--$r_d$ &
       $H_0$--$\Omega_m$ & $H_0$--$M$ & $H_0$--$r_d$ \\
\hline
1 & 0.032 & $-0.123$ & $0.984$ & $\cdots$ & $-33.0$ & $96.0$ & $\cdots$  \\
2 & 0.150 & $-0.693$ & $0.963$ & $\cdots$ & $-93.0$ & $84.2$ & $\cdots$  \\
3 & 0.300 & $-0.557$ & $0.975$ & $-0.944$ & $-28.9$ & $93.4$ & $-64.2$ \\
4 & 0.510 & $-0.631$ & $0.943$ & $-0.890$ & $-38.8$ & $91.8$ & $-65.4$ \\
5 & 0.710 & $-0.590$ & $0.939$ & $-0.906$ & $-22.8$ & $92.0$ & $-65.7$ \\
6 & 0.930 & $-0.404$ & $0.922$ & $-0.953$ & $-6.1$  & $86.1$ & $-59.5$ \\
7 & 1.380 & $-0.645$ & $0.818$ & $-0.886$ & $-14.1$ & $77.5$ & $-59.4$ \\
8 & 1.940 & $-0.347$ & $0.908$ & $-0.945$ & $-3.8$  & $95.8$ & $-64.2$ \\
\end{tabular}
\end{ruledtabular}
\end{table*}

\bibliography{reference}

\begin{thebibliography}{49}%
\makeatletter
\providecommand \@ifxundefined [1]{%
 \@ifx{#1\undefined}
}%
\providecommand \@ifnum [1]{%
 \ifnum #1\expandafter \@firstoftwo
 \else \expandafter \@secondoftwo
 \fi
}%
\providecommand \@ifx [1]{%
 \ifx #1\expandafter \@firstoftwo
 \else \expandafter \@secondoftwo
 \fi
}%
\providecommand \natexlab [1]{#1}%
\providecommand \enquote  [1]{``#1''}%
\providecommand \bibnamefont  [1]{#1}%
\providecommand \bibfnamefont [1]{#1}%
\providecommand \citenamefont [1]{#1}%
\providecommand \href@noop [0]{\@secondoftwo}%
\providecommand \href [0]{\begingroup \@sanitize@url \@href}%
\providecommand \@href[1]{\@@startlink{#1}\@@href}%
\providecommand \@@href[1]{\endgroup#1\@@endlink}%
\providecommand \@sanitize@url [0]{\catcode `\\12\catcode `\$12\catcode `\&12\catcode `\#12\catcode `\^12\catcode `\_12\catcode `\%12\relax}%
\providecommand \@@startlink[1]{}%
\providecommand \@@endlink[0]{}%
\providecommand \url  [0]{\begingroup\@sanitize@url \@url }%
\providecommand \@url [1]{\endgroup\@href {#1}{\urlprefix }}%
\providecommand \urlprefix  [0]{URL }%
\providecommand \Eprint [0]{\href }%
\providecommand \doibase [0]{https://doi.org/}%
\providecommand \selectlanguage [0]{\@gobble}%
\providecommand \bibinfo  [0]{\@secondoftwo}%
\providecommand \bibfield  [0]{\@secondoftwo}%
\providecommand \translation [1]{[#1]}%
\providecommand \BibitemOpen [0]{}%
\providecommand \bibitemStop [0]{}%
\providecommand \bibitemNoStop [0]{.\EOS\space}%
\providecommand \EOS [0]{\spacefactor3000\relax}%
\providecommand \BibitemShut  [1]{\csname bibitem#1\endcsname}%
\let\auto@bib@innerbib\@empty
\bibitem [{\citenamefont {Collaboration}\ \emph {et~al.}(2020)\citenamefont {Collaboration}, \citenamefont {Aghanim}, \citenamefont {Akrami}, \citenamefont {Ashdown}, \citenamefont {Aumont} \emph {et~al.}}]{Planck2018}%
  \BibitemOpen
  \bibfield  {author} {\bibinfo {author} {\bibfnamefont {P.}~\bibnamefont {Collaboration}}, \bibinfo {author} {\bibfnamefont {N.}~\bibnamefont {Aghanim}}, \bibinfo {author} {\bibfnamefont {Y.}~\bibnamefont {Akrami}}, \bibinfo {author} {\bibfnamefont {M.}~\bibnamefont {Ashdown}}, \bibinfo {author} {\bibfnamefont {J.}~\bibnamefont {Aumont}}, \emph {et~al.},\ }\bibfield  {title} {\bibinfo {title} {Planck 2018 results. vi. cosmological parameters},\ }\href {https://doi.org/10.1051/0004-6361/201833910} {\bibfield  {journal} {\bibinfo  {journal} {Astronomy \& Astrophysics}\ }\textbf {\bibinfo {volume} {641}},\ \bibinfo {pages} {A6} (\bibinfo {year} {2020})},\ \Eprint {https://arxiv.org/abs/1807.06209} {arXiv:1807.06209 [astro-ph.CO]} \BibitemShut {NoStop}%
\bibitem [{\citenamefont {Eisenstein}\ \emph {et~al.}(2005)\citenamefont {Eisenstein}, \citenamefont {Zehavi}, \citenamefont {Hogg}, \citenamefont {Scoccimarro}, \citenamefont {Blanton}, \citenamefont {Nichol}, \citenamefont {Scranton}, \citenamefont {Seo}, \citenamefont {Tegmark}, \citenamefont {Zheng} \emph {et~al.}}]{Eisenstein2005}%
  \BibitemOpen
  \bibfield  {author} {\bibinfo {author} {\bibfnamefont {D.~J.}\ \bibnamefont {Eisenstein}}, \bibinfo {author} {\bibfnamefont {I.}~\bibnamefont {Zehavi}}, \bibinfo {author} {\bibfnamefont {D.~W.}\ \bibnamefont {Hogg}}, \bibinfo {author} {\bibfnamefont {R.}~\bibnamefont {Scoccimarro}}, \bibinfo {author} {\bibfnamefont {M.~R.}\ \bibnamefont {Blanton}}, \bibinfo {author} {\bibfnamefont {R.~C.}\ \bibnamefont {Nichol}}, \bibinfo {author} {\bibfnamefont {R.}~\bibnamefont {Scranton}}, \bibinfo {author} {\bibfnamefont {H.-J.}\ \bibnamefont {Seo}}, \bibinfo {author} {\bibfnamefont {M.}~\bibnamefont {Tegmark}}, \bibinfo {author} {\bibfnamefont {Z.}~\bibnamefont {Zheng}}, \emph {et~al.},\ }\bibfield  {title} {\bibinfo {title} {Detection of the baryon acoustic peak in the large-scale correlation function of sdss luminous red galaxies},\ }\href {https://doi.org/10.1086/466512} {\bibfield  {journal} {\bibinfo  {journal} {The Astrophysical Journal}\ }\textbf {\bibinfo {volume} {633}},\ \bibinfo {pages} {560} (\bibinfo
  {year} {2005})},\ \Eprint {https://arxiv.org/abs/astro-ph/0501171} {arXiv:astro-ph/0501171 [astro-ph]} \BibitemShut {NoStop}%
\bibitem [{\citenamefont {Alam}\ \emph {et~al.}(2017)\citenamefont {Alam}, \citenamefont {Ata}, \citenamefont {Bailey}, \citenamefont {Beutler}, \citenamefont {Bizyaev} \emph {et~al.}}]{Alam2017}%
  \BibitemOpen
  \bibfield  {author} {\bibinfo {author} {\bibfnamefont {S.}~\bibnamefont {Alam}}, \bibinfo {author} {\bibfnamefont {M.}~\bibnamefont {Ata}}, \bibinfo {author} {\bibfnamefont {S.}~\bibnamefont {Bailey}}, \bibinfo {author} {\bibfnamefont {F.}~\bibnamefont {Beutler}}, \bibinfo {author} {\bibfnamefont {D.}~\bibnamefont {Bizyaev}}, \emph {et~al.},\ }\bibfield  {title} {\bibinfo {title} {The clustering of galaxies in the completed sdss-iii baryon oscillation spectroscopic survey: cosmological analysis of the dr12 galaxy sample},\ }\href {https://doi.org/10.1093/mnras/stx721} {\bibfield  {journal} {\bibinfo  {journal} {Monthly Notices of the Royal Astronomical Society}\ }\textbf {\bibinfo {volume} {470}},\ \bibinfo {pages} {2617} (\bibinfo {year} {2017})},\ \Eprint {https://arxiv.org/abs/1607.03155} {arXiv:1607.03155 [astro-ph.CO]} \BibitemShut {NoStop}%
\bibitem [{\citenamefont {Suzuki}\ \emph {et~al.}(2012)\citenamefont {Suzuki}, \citenamefont {Rubin}, \citenamefont {Lidman}, \citenamefont {Aldering}, \citenamefont {Amanullah}, \citenamefont {Barbary}, \citenamefont {Barrientos} \emph {et~al.}}]{Suzuki2012}%
  \BibitemOpen
  \bibfield  {author} {\bibinfo {author} {\bibfnamefont {N.}~\bibnamefont {Suzuki}}, \bibinfo {author} {\bibfnamefont {D.}~\bibnamefont {Rubin}}, \bibinfo {author} {\bibfnamefont {C.}~\bibnamefont {Lidman}}, \bibinfo {author} {\bibfnamefont {G.}~\bibnamefont {Aldering}}, \bibinfo {author} {\bibfnamefont {R.}~\bibnamefont {Amanullah}}, \bibinfo {author} {\bibfnamefont {K.}~\bibnamefont {Barbary}}, \bibinfo {author} {\bibfnamefont {L.~F.}\ \bibnamefont {Barrientos}}, \emph {et~al.},\ }\bibfield  {title} {\bibinfo {title} {The {Hubble Space Telescope} cluster supernova survey. v. improving the dark-energy constraints above $z > 1$ and building an early-type-hosted supernova sample},\ }\href {https://doi.org/10.1088/0004-637X/746/1/85} {\bibfield  {journal} {\bibinfo  {journal} {The Astrophysical Journal}\ }\textbf {\bibinfo {volume} {746}},\ \bibinfo {pages} {85} (\bibinfo {year} {2012})},\ \Eprint {https://arxiv.org/abs/1105.3470} {arXiv:1105.3470 [astro-ph.CO]} \BibitemShut {NoStop}%
\bibitem [{\citenamefont {Betoule}\ \emph {et~al.}(2014)\citenamefont {Betoule}, \citenamefont {Kessler}, \citenamefont {Guy}, \citenamefont {Mosher}, \citenamefont {Hardin} \emph {et~al.}}]{Betoule2014}%
  \BibitemOpen
  \bibfield  {author} {\bibinfo {author} {\bibfnamefont {M.}~\bibnamefont {Betoule}}, \bibinfo {author} {\bibfnamefont {R.}~\bibnamefont {Kessler}}, \bibinfo {author} {\bibfnamefont {J.}~\bibnamefont {Guy}}, \bibinfo {author} {\bibfnamefont {J.}~\bibnamefont {Mosher}}, \bibinfo {author} {\bibfnamefont {D.}~\bibnamefont {Hardin}}, \emph {et~al.},\ }\bibfield  {title} {\bibinfo {title} {Improved cosmological constraints from a joint analysis of the {SDSS-II} and {SNLS} supernova samples},\ }\href {https://doi.org/10.1051/0004-6361/201423413} {\bibfield  {journal} {\bibinfo  {journal} {Astronomy \& Astrophysics}\ }\textbf {\bibinfo {volume} {568}},\ \bibinfo {pages} {A22} (\bibinfo {year} {2014})},\ \Eprint {https://arxiv.org/abs/1401.4064} {arXiv:1401.4064 [astro-ph.CO]} \BibitemShut {NoStop}%
\bibitem [{\citenamefont {Brout}\ \emph {et~al.}(2022)\citenamefont {Brout}, \citenamefont {Scolnic}, \citenamefont {Popovic}, \citenamefont {Riess}, \citenamefont {Carr}, \citenamefont {Zuntz}, \citenamefont {Kessler}, \citenamefont {Davis}, \citenamefont {Hinton}, \citenamefont {Jones} \emph {et~al.}}]{Brout2022}%
  \BibitemOpen
  \bibfield  {author} {\bibinfo {author} {\bibfnamefont {D.}~\bibnamefont {Brout}}, \bibinfo {author} {\bibfnamefont {D.}~\bibnamefont {Scolnic}}, \bibinfo {author} {\bibfnamefont {B.}~\bibnamefont {Popovic}}, \bibinfo {author} {\bibfnamefont {A.~G.}\ \bibnamefont {Riess}}, \bibinfo {author} {\bibfnamefont {A.}~\bibnamefont {Carr}}, \bibinfo {author} {\bibfnamefont {J.}~\bibnamefont {Zuntz}}, \bibinfo {author} {\bibfnamefont {R.}~\bibnamefont {Kessler}}, \bibinfo {author} {\bibfnamefont {T.~M.}\ \bibnamefont {Davis}}, \bibinfo {author} {\bibfnamefont {S.}~\bibnamefont {Hinton}}, \bibinfo {author} {\bibfnamefont {D.~O.}\ \bibnamefont {Jones}}, \emph {et~al.},\ }\bibfield  {title} {\bibinfo {title} {The {Pantheon+} analysis: Cosmological constraints},\ }\href {https://doi.org/10.3847/1538-4357/ac8e04} {\bibfield  {journal} {\bibinfo  {journal} {The Astrophysical Journal}\ }\textbf {\bibinfo {volume} {938}},\ \bibinfo {pages} {110} (\bibinfo {year} {2022})},\ \Eprint {https://arxiv.org/abs/2202.04077}
  {arXiv:2202.04077 [astro-ph.CO]} \BibitemShut {NoStop}%
\bibitem [{\citenamefont {Riess}\ \emph {et~al.}(2022)\citenamefont {Riess}, \citenamefont {Yuan}, \citenamefont {Macri}, \citenamefont {Scolnic}, \citenamefont {Brout} \emph {et~al.}}]{Riess2022}%
  \BibitemOpen
  \bibfield  {author} {\bibinfo {author} {\bibfnamefont {A.~G.}\ \bibnamefont {Riess}}, \bibinfo {author} {\bibfnamefont {W.}~\bibnamefont {Yuan}}, \bibinfo {author} {\bibfnamefont {L.~M.}\ \bibnamefont {Macri}}, \bibinfo {author} {\bibfnamefont {D.}~\bibnamefont {Scolnic}}, \bibinfo {author} {\bibfnamefont {D.}~\bibnamefont {Brout}}, \emph {et~al.},\ }\bibfield  {title} {\bibinfo {title} {A comprehensive measurement of the local value of the hubble constant with 1 km s$^{-1}$ mpc$^{-1}$ uncertainty from the hubble space telescope and the {SH0ES} team},\ }\href {https://doi.org/10.3847/2041-8213/ac5c5b} {\bibfield  {journal} {\bibinfo  {journal} {The Astrophysical Journal Letters}\ }\textbf {\bibinfo {volume} {934}},\ \bibinfo {pages} {L7} (\bibinfo {year} {2022})},\ \Eprint {https://arxiv.org/abs/2112.04510} {arXiv:2112.04510 [astro-ph.CO]} \BibitemShut {NoStop}%
\bibitem [{\citenamefont {Bonvin}\ \emph {et~al.}(2017)\citenamefont {Bonvin}, \citenamefont {Courbin}, \citenamefont {Suyu}, \citenamefont {Marshall}, \citenamefont {Rusu} \emph {et~al.}}]{Bonvin2017}%
  \BibitemOpen
  \bibfield  {author} {\bibinfo {author} {\bibfnamefont {V.}~\bibnamefont {Bonvin}}, \bibinfo {author} {\bibfnamefont {F.}~\bibnamefont {Courbin}}, \bibinfo {author} {\bibfnamefont {S.~H.}\ \bibnamefont {Suyu}}, \bibinfo {author} {\bibfnamefont {P.~J.}\ \bibnamefont {Marshall}}, \bibinfo {author} {\bibfnamefont {C.~E.}\ \bibnamefont {Rusu}}, \emph {et~al.},\ }\bibfield  {title} {\bibinfo {title} {H0licow -- v. new cosmograil time delays of he 0435-1223: H0 to 3.8 per cent precision from strong lensing in a flat {$\Lambda$}cdm model},\ }\href {https://doi.org/10.1093/mnras/stw3006} {\bibfield  {journal} {\bibinfo  {journal} {Monthly Notices of the Royal Astronomical Society}\ }\textbf {\bibinfo {volume} {465}},\ \bibinfo {pages} {4914} (\bibinfo {year} {2017})},\ \Eprint {https://arxiv.org/abs/1607.01790} {arXiv:1607.01790 [astro-ph.CO]} \BibitemShut {NoStop}%
\bibitem [{\citenamefont {Wong}\ \emph {et~al.}(2020)\citenamefont {Wong}, \citenamefont {Suyu}, \citenamefont {Chen}, \citenamefont {Rusu}, \citenamefont {Millon} \emph {et~al.}}]{Wong2019}%
  \BibitemOpen
  \bibfield  {author} {\bibinfo {author} {\bibfnamefont {K.~C.}\ \bibnamefont {Wong}}, \bibinfo {author} {\bibfnamefont {S.~H.}\ \bibnamefont {Suyu}}, \bibinfo {author} {\bibfnamefont {G.~C.-F.}\ \bibnamefont {Chen}}, \bibinfo {author} {\bibfnamefont {C.~E.}\ \bibnamefont {Rusu}}, \bibinfo {author} {\bibfnamefont {M.}~\bibnamefont {Millon}}, \emph {et~al.},\ }\bibfield  {title} {\bibinfo {title} {H0licow -- xiii. a 2.4 per cent measurement of h0 from lensed quasars: 5.3$\sigma$ tension between early- and late-universe probes},\ }\href {https://doi.org/10.1093/mnras/stz3094} {\bibfield  {journal} {\bibinfo  {journal} {Monthly Notices of the Royal Astronomical Society}\ }\textbf {\bibinfo {volume} {498}},\ \bibinfo {pages} {1420} (\bibinfo {year} {2020})},\ \Eprint {https://arxiv.org/abs/1907.04869} {arXiv:1907.04869 [astro-ph.CO]} \BibitemShut {NoStop}%
\bibitem [{\citenamefont {Krishnan}\ \emph {et~al.}(2020)\citenamefont {Krishnan}, \citenamefont {Colg{\'a}in}, \citenamefont {Ruchika}, \citenamefont {Sen}, \citenamefont {Sheikh-Jabbari},\ and\ \citenamefont {Yang}}]{Krishnan2020}%
  \BibitemOpen
  \bibfield  {author} {\bibinfo {author} {\bibfnamefont {C.}~\bibnamefont {Krishnan}}, \bibinfo {author} {\bibfnamefont {E.~{\'O}.}\ \bibnamefont {Colg{\'a}in}}, \bibinfo {author} {\bibnamefont {Ruchika}}, \bibinfo {author} {\bibfnamefont {A.~A.}\ \bibnamefont {Sen}}, \bibinfo {author} {\bibfnamefont {M.~M.}\ \bibnamefont {Sheikh-Jabbari}},\ and\ \bibinfo {author} {\bibfnamefont {T.}~\bibnamefont {Yang}},\ }\bibfield  {title} {\bibinfo {title} {Is there an early universe solution to hubble tension?},\ }\href {https://doi.org/10.1103/PhysRevD.102.103525} {\bibfield  {journal} {\bibinfo  {journal} {Physical Review D}\ }\textbf {\bibinfo {volume} {102}},\ \bibinfo {pages} {103525} (\bibinfo {year} {2020})},\ \Eprint {https://arxiv.org/abs/2002.06044} {arXiv:2002.06044 [astro-ph.CO]} \BibitemShut {NoStop}%
\bibitem [{\citenamefont {Dainotti}\ \emph {et~al.}(2024)\citenamefont {Dainotti}, \citenamefont {De~Simone}, \citenamefont {Bogdan},\ and\ \citenamefont {Montani}}]{Dainotti2024}%
  \BibitemOpen
  \bibfield  {author} {\bibinfo {author} {\bibfnamefont {M.~G.}\ \bibnamefont {Dainotti}}, \bibinfo {author} {\bibfnamefont {B.}~\bibnamefont {De~Simone}}, \bibinfo {author} {\bibfnamefont {M.}~\bibnamefont {Bogdan}},\ and\ \bibinfo {author} {\bibfnamefont {G.}~\bibnamefont {Montani}},\ }\bibfield  {title} {\bibinfo {title} {Shedding new light on the hubble constant tension through supernovae ia},\ }\href {https://doi.org/10.22323/1.447.0068} {\bibfield  {journal} {\bibinfo  {journal} {Proceedings of Science (PoS), MULTIF2023}\ }\textbf {\bibinfo {volume} {MULTIF2023}},\ \bibinfo {pages} {068} (\bibinfo {year} {2024})},\ \Eprint {https://arxiv.org/abs/2311.15188} {arXiv:2311.15188 [astro-ph.CO]} \BibitemShut {NoStop}%
\bibitem [{\citenamefont {Lopez-Hernandez}\ and\ \citenamefont {De-Santiago}(2025)}]{LopezHernandez2025}%
  \BibitemOpen
  \bibfield  {author} {\bibinfo {author} {\bibfnamefont {M.}~\bibnamefont {Lopez-Hernandez}}\ and\ \bibinfo {author} {\bibfnamefont {J.}~\bibnamefont {De-Santiago}},\ }\bibfield  {title} {\bibinfo {title} {Is there a dynamical tendency in h0 with late time measurements?},\ }\href {https://doi.org/10.1088/1475-7516/2025/03/026} {\bibfield  {journal} {\bibinfo  {journal} {JCAP}\ }\textbf {\bibinfo {volume} {2025}}\bibfield  {number} {\bibinfo  {number} { (03)},\ \bibinfo {pages} {026}},\ }\Eprint {https://arxiv.org/abs/2411.00095} {arXiv:2411.00095 [astro-ph.CO]} \BibitemShut {NoStop}%
\bibitem [{\citenamefont {Dainotti}\ \emph {et~al.}(2025)\citenamefont {Dainotti}, \citenamefont {De~Simone}, \citenamefont {Garg}, \citenamefont {Kohri}, \citenamefont {Bashyal}, \citenamefont {Aich}, \citenamefont {Mondal}, \citenamefont {Nagataki}, \citenamefont {Montani}, \citenamefont {Jareen}, \citenamefont {Jabir}, \citenamefont {Khanjani}, \citenamefont {Bogdan}, \citenamefont {Fraija}, \citenamefont {do~E.~S.~Pedreira}, \citenamefont {Dejrah}, \citenamefont {Singh}, \citenamefont {Parakh}, \citenamefont {Mandal}, \citenamefont {Jarial}, \citenamefont {Lambiase},\ and\ \citenamefont {Sarkar}}]{Dainotti2025}%
  \BibitemOpen
  \bibfield  {author} {\bibinfo {author} {\bibfnamefont {M.~G.}\ \bibnamefont {Dainotti}}, \bibinfo {author} {\bibfnamefont {B.}~\bibnamefont {De~Simone}}, \bibinfo {author} {\bibfnamefont {A.}~\bibnamefont {Garg}}, \bibinfo {author} {\bibfnamefont {K.}~\bibnamefont {Kohri}}, \bibinfo {author} {\bibfnamefont {A.}~\bibnamefont {Bashyal}}, \bibinfo {author} {\bibfnamefont {A.}~\bibnamefont {Aich}}, \bibinfo {author} {\bibfnamefont {A.}~\bibnamefont {Mondal}}, \bibinfo {author} {\bibfnamefont {S.}~\bibnamefont {Nagataki}}, \bibinfo {author} {\bibfnamefont {G.}~\bibnamefont {Montani}}, \bibinfo {author} {\bibfnamefont {T.}~\bibnamefont {Jareen}}, \bibinfo {author} {\bibfnamefont {V.~M.}\ \bibnamefont {Jabir}}, \bibinfo {author} {\bibfnamefont {S.}~\bibnamefont {Khanjani}}, \bibinfo {author} {\bibfnamefont {M.}~\bibnamefont {Bogdan}}, \bibinfo {author} {\bibfnamefont {N.}~\bibnamefont {Fraija}}, \bibinfo {author} {\bibfnamefont {A.~C.~C.}\ \bibnamefont {do~E.~S.~Pedreira}}, \bibinfo {author} {\bibfnamefont
  {R.~H.}\ \bibnamefont {Dejrah}}, \bibinfo {author} {\bibfnamefont {A.}~\bibnamefont {Singh}}, \bibinfo {author} {\bibfnamefont {M.}~\bibnamefont {Parakh}}, \bibinfo {author} {\bibfnamefont {R.}~\bibnamefont {Mandal}}, \bibinfo {author} {\bibfnamefont {K.}~\bibnamefont {Jarial}}, \bibinfo {author} {\bibfnamefont {G.}~\bibnamefont {Lambiase}},\ and\ \bibinfo {author} {\bibfnamefont {H.}~\bibnamefont {Sarkar}},\ }\bibfield  {title} {\bibinfo {title} {A new master supernovae ia sample and the investigation of the hubble tension},\ }\href {https://doi.org/10.1016/j.jheap.2025.100405} {\bibfield  {journal} {\bibinfo  {journal} {Journal of High Energy Astrophysics}\ }\textbf {\bibinfo {volume} {48}},\ \bibinfo {pages} {100405} (\bibinfo {year} {2025})},\ \Eprint {https://arxiv.org/abs/2501.11772} {arXiv:2501.11772 [astro-ph.CO]} \BibitemShut {NoStop}%
\bibitem [{\citenamefont {Scolnic}\ \emph {et~al.}(2022)\citenamefont {Scolnic}, \citenamefont {Brout}, \citenamefont {Carr} \emph {et~al.}}]{Scolnic2022}%
  \BibitemOpen
  \bibfield  {author} {\bibinfo {author} {\bibfnamefont {D.}~\bibnamefont {Scolnic}}, \bibinfo {author} {\bibfnamefont {D.}~\bibnamefont {Brout}}, \bibinfo {author} {\bibfnamefont {A.}~\bibnamefont {Carr}}, \emph {et~al.},\ }\bibfield  {title} {\bibinfo {title} {The pantheon+ analysis: The full data set and light-curve release},\ }\href {https://doi.org/10.3847/1538-4357/ac8b7a} {\bibfield  {journal} {\bibinfo  {journal} {The Astrophysical Journal}\ }\textbf {\bibinfo {volume} {938}},\ \bibinfo {pages} {113} (\bibinfo {year} {2022})}\BibitemShut {NoStop}%
\bibitem [{\citenamefont {Malekjani}\ \emph {et~al.}(2024)\citenamefont {Malekjani}, \citenamefont {Mc~Conville}, \citenamefont {Ó~Colgáin}, \citenamefont {Pourojaghi},\ and\ \citenamefont {Sheikh-Jabbari}}]{Malekjani2024}%
  \BibitemOpen
  \bibfield  {author} {\bibinfo {author} {\bibfnamefont {M.}~\bibnamefont {Malekjani}}, \bibinfo {author} {\bibfnamefont {R.}~\bibnamefont {Mc~Conville}}, \bibinfo {author} {\bibfnamefont {E.}~\bibnamefont {Ó~Colgáin}}, \bibinfo {author} {\bibfnamefont {S.}~\bibnamefont {Pourojaghi}},\ and\ \bibinfo {author} {\bibfnamefont {M.~M.}\ \bibnamefont {Sheikh-Jabbari}},\ }\bibfield  {title} {\bibinfo {title} {On redshift evolution and negative dark energy density in pantheon+ supernovae},\ }\href {https://doi.org/10.1140/epjc/s10052-024-12667-z} {\bibfield  {journal} {\bibinfo  {journal} {The European Physical Journal C}\ }\textbf {\bibinfo {volume} {84}},\ \bibinfo {pages} {317} (\bibinfo {year} {2024})}\BibitemShut {NoStop}%
\bibitem [{\citenamefont {Abdul~Karim}\ \emph {et~al.}(2025)\citenamefont {Abdul~Karim}, \citenamefont {Alam}, \citenamefont {Aviles}, \citenamefont {Dawson}, \citenamefont {Eisenstein}, \citenamefont {Ferraro}, \citenamefont {Gil-Mar{\'\i}n}, \citenamefont {Levi}, \citenamefont {Nadathur}, \citenamefont {Percival}, \citenamefont {Ross}, \citenamefont {Schlegel}, \citenamefont {Seo}, \citenamefont {White}, \citenamefont {Zhao}, \citenamefont {{DESI Collaboration}},\ and\ \citenamefont {et~al.}}]{DESI_DR2}%
  \BibitemOpen
  \bibfield  {author} {\bibinfo {author} {\bibfnamefont {M.}~\bibnamefont {Abdul~Karim}}, \bibinfo {author} {\bibfnamefont {S.}~\bibnamefont {Alam}}, \bibinfo {author} {\bibfnamefont {A.}~\bibnamefont {Aviles}}, \bibinfo {author} {\bibfnamefont {K.~S.}\ \bibnamefont {Dawson}}, \bibinfo {author} {\bibfnamefont {D.~J.}\ \bibnamefont {Eisenstein}}, \bibinfo {author} {\bibfnamefont {S.}~\bibnamefont {Ferraro}}, \bibinfo {author} {\bibfnamefont {H.}~\bibnamefont {Gil-Mar{\'\i}n}}, \bibinfo {author} {\bibfnamefont {M.~E.}\ \bibnamefont {Levi}}, \bibinfo {author} {\bibfnamefont {S.}~\bibnamefont {Nadathur}}, \bibinfo {author} {\bibfnamefont {W.~J.}\ \bibnamefont {Percival}}, \bibinfo {author} {\bibfnamefont {A.~J.}\ \bibnamefont {Ross}}, \bibinfo {author} {\bibfnamefont {D.~J.}\ \bibnamefont {Schlegel}}, \bibinfo {author} {\bibfnamefont {H.-J.}\ \bibnamefont {Seo}}, \bibinfo {author} {\bibfnamefont {M.}~\bibnamefont {White}}, \bibinfo {author} {\bibfnamefont {G.-B.}\ \bibnamefont {Zhao}}, \bibinfo {author}
  {\bibnamefont {{DESI Collaboration}}},\ and\ \bibinfo {author} {\bibnamefont {et~al.}},\ }\bibfield  {title} {\bibinfo {title} {Desi dr2 results. ii. measurements of baryon acoustic oscillations and cosmological constraints},\ }\href {https://doi.org/10.1103/tr6y-kpc6} {\bibfield  {journal} {\bibinfo  {journal} {Physical Review D}\ }\textbf {\bibinfo {volume} {112}},\ \bibinfo {pages} {083515} (\bibinfo {year} {2025})}\BibitemShut {NoStop}%
\bibitem [{\citenamefont {Jimenez}\ \emph {et~al.}(2003)\citenamefont {Jimenez}, \citenamefont {Verde}, \citenamefont {Treu},\ and\ \citenamefont {Stern}}]{Jimenez2003}%
  \BibitemOpen
  \bibfield  {author} {\bibinfo {author} {\bibfnamefont {R.}~\bibnamefont {Jimenez}}, \bibinfo {author} {\bibfnamefont {L.}~\bibnamefont {Verde}}, \bibinfo {author} {\bibfnamefont {T.}~\bibnamefont {Treu}},\ and\ \bibinfo {author} {\bibfnamefont {D.}~\bibnamefont {Stern}},\ }\bibfield  {title} {\bibinfo {title} {Constraints on the equation of state of dark energy and the hubble constant from stellar ages and the cosmic microwave background},\ }\href {https://doi.org/10.1086/376595} {\bibfield  {journal} {\bibinfo  {journal} {Astrophys. J.}\ }\textbf {\bibinfo {volume} {593}},\ \bibinfo {pages} {622} (\bibinfo {year} {2003})}\BibitemShut {NoStop}%
\bibitem [{\citenamefont {Niu}\ \emph {et~al.}(2025)\citenamefont {Niu}, \citenamefont {He},\ and\ \citenamefont {Zhang}}]{Niu2025}%
  \BibitemOpen
  \bibfield  {author} {\bibinfo {author} {\bibfnamefont {J.}~\bibnamefont {Niu}}, \bibinfo {author} {\bibfnamefont {P.}~\bibnamefont {He}},\ and\ \bibinfo {author} {\bibfnamefont {T.-J.}\ \bibnamefont {Zhang}},\ }\href {https://arxiv.org/abs/2502.11443} {\bibinfo {title} {Constraining the hubble constant with a simulated full covariance matrix using neural networks}} (\bibinfo {year} {2025}),\ \Eprint {https://arxiv.org/abs/2502.11443} {arXiv:2502.11443 [astro-ph.CO]} \BibitemShut {NoStop}%
\bibitem [{\citenamefont {Pesce}\ \emph {et~al.}(2020)\citenamefont {Pesce}, \citenamefont {Braatz}, \citenamefont {Reid}, \citenamefont {Riess}, \citenamefont {Scolnic}, \citenamefont {Condon}, \citenamefont {Gao}, \citenamefont {Henkel}, \citenamefont {Impellizzeri}, \citenamefont {Kuo},\ and\ \citenamefont {Lo}}]{Pesce2020}%
  \BibitemOpen
  \bibfield  {author} {\bibinfo {author} {\bibfnamefont {D.~W.}\ \bibnamefont {Pesce}}, \bibinfo {author} {\bibfnamefont {J.~A.}\ \bibnamefont {Braatz}}, \bibinfo {author} {\bibfnamefont {M.~J.}\ \bibnamefont {Reid}}, \bibinfo {author} {\bibfnamefont {A.~G.}\ \bibnamefont {Riess}}, \bibinfo {author} {\bibfnamefont {D.}~\bibnamefont {Scolnic}}, \bibinfo {author} {\bibfnamefont {J.~J.}\ \bibnamefont {Condon}}, \bibinfo {author} {\bibfnamefont {F.}~\bibnamefont {Gao}}, \bibinfo {author} {\bibfnamefont {C.}~\bibnamefont {Henkel}}, \bibinfo {author} {\bibfnamefont {C.~M.~V.}\ \bibnamefont {Impellizzeri}}, \bibinfo {author} {\bibfnamefont {C.~Y.}\ \bibnamefont {Kuo}},\ and\ \bibinfo {author} {\bibfnamefont {K.~Y.}\ \bibnamefont {Lo}},\ }\bibfield  {title} {\bibinfo {title} {The megamaser cosmology project. xiii. combined hubble constant constraints},\ }\href {https://doi.org/10.3847/2041-8213/ab75f0} {\bibfield  {journal} {\bibinfo  {journal} {Astrophys. J. Lett.}\ }\textbf {\bibinfo {volume} {891}},\ \bibinfo
  {pages} {L1} (\bibinfo {year} {2020})},\ \Eprint {https://arxiv.org/abs/2001.09213} {arXiv:2001.09213 [astro-ph.CO]} \BibitemShut {NoStop}%
\bibitem [{\citenamefont {Zhang}\ \emph {et~al.}(2014)\citenamefont {Zhang}, \citenamefont {Zhang}, \citenamefont {Yuan}, \citenamefont {Liu}, \citenamefont {Zhang},\ and\ \citenamefont {Sun}}]{Zhang2014}%
  \BibitemOpen
  \bibfield  {author} {\bibinfo {author} {\bibfnamefont {C.}~\bibnamefont {Zhang}}, \bibinfo {author} {\bibfnamefont {H.}~\bibnamefont {Zhang}}, \bibinfo {author} {\bibfnamefont {S.}~\bibnamefont {Yuan}}, \bibinfo {author} {\bibfnamefont {S.}~\bibnamefont {Liu}}, \bibinfo {author} {\bibfnamefont {T.-J.}\ \bibnamefont {Zhang}},\ and\ \bibinfo {author} {\bibfnamefont {Y.-C.}\ \bibnamefont {Sun}},\ }\bibfield  {title} {\bibinfo {title} {Four new observational h(z) data from luminous red galaxies in the sloan digital sky survey data release seven},\ }\href {https://doi.org/10.1088/1674-4527/14/10/002} {\bibfield  {journal} {\bibinfo  {journal} {Research in Astronomy and Astrophysics}\ }\textbf {\bibinfo {volume} {14}},\ \bibinfo {pages} {1221} (\bibinfo {year} {2014})},\ \Eprint {https://arxiv.org/abs/1207.4541} {arXiv:1207.4541 [astro-ph.CO]} \BibitemShut {NoStop}%
\bibitem [{\citenamefont {Loubser}\ \emph {et~al.}(2025)\citenamefont {Loubser}, \citenamefont {Alabi}, \citenamefont {Hilton}, \citenamefont {Ma}, \citenamefont {Tang}, \citenamefont {Hatamkhani}, \citenamefont {Cress}, \citenamefont {Skelton},\ and\ \citenamefont {Nkosi}}]{Loubser2025}%
  \BibitemOpen
  \bibfield  {author} {\bibinfo {author} {\bibfnamefont {S.~I.}\ \bibnamefont {Loubser}}, \bibinfo {author} {\bibfnamefont {A.~B.}\ \bibnamefont {Alabi}}, \bibinfo {author} {\bibfnamefont {M.}~\bibnamefont {Hilton}}, \bibinfo {author} {\bibfnamefont {Y.-Z.}\ \bibnamefont {Ma}}, \bibinfo {author} {\bibfnamefont {X.}~\bibnamefont {Tang}}, \bibinfo {author} {\bibfnamefont {N.}~\bibnamefont {Hatamkhani}}, \bibinfo {author} {\bibfnamefont {C.}~\bibnamefont {Cress}}, \bibinfo {author} {\bibfnamefont {R.~E.}\ \bibnamefont {Skelton}},\ and\ \bibinfo {author} {\bibfnamefont {S.~A.}\ \bibnamefont {Nkosi}},\ }\href {https://arxiv.org/abs/2506.03836} {\bibinfo {title} {An independent estimate of $h(z)$ at $z = 0.5$ from the stellar ages of brightest cluster galaxies}} (\bibinfo {year} {2025}),\ \Eprint {https://arxiv.org/abs/2506.03836} {arXiv:2506.03836 [astro-ph.CO]} \BibitemShut {NoStop}%
\bibitem [{\citenamefont {Moresco}\ \emph {et~al.}(2012)\citenamefont {Moresco}, \citenamefont {Cimatti}, \citenamefont {Jimenez}, \citenamefont {Pozzetti}, \citenamefont {Zamorani}, \citenamefont {Bolzonella}, \citenamefont {Dunlop}, \citenamefont {Lamareille}, \citenamefont {Mignoli}, \citenamefont {Pearce}, \citenamefont {Rosati}, \citenamefont {Stern}, \citenamefont {Verde}, \citenamefont {Zucca}, \citenamefont {Carollo}, \citenamefont {Contini}, \citenamefont {Kneib}, \citenamefont {Le~F{\`e}vre}, \citenamefont {Lilly}, \citenamefont {Mainieri} \emph {et~al.}}]{Moresco2012}%
  \BibitemOpen
  \bibfield  {author} {\bibinfo {author} {\bibfnamefont {M.}~\bibnamefont {Moresco}}, \bibinfo {author} {\bibfnamefont {A.}~\bibnamefont {Cimatti}}, \bibinfo {author} {\bibfnamefont {R.}~\bibnamefont {Jimenez}}, \bibinfo {author} {\bibfnamefont {L.}~\bibnamefont {Pozzetti}}, \bibinfo {author} {\bibfnamefont {G.}~\bibnamefont {Zamorani}}, \bibinfo {author} {\bibfnamefont {M.}~\bibnamefont {Bolzonella}}, \bibinfo {author} {\bibfnamefont {J.}~\bibnamefont {Dunlop}}, \bibinfo {author} {\bibfnamefont {F.}~\bibnamefont {Lamareille}}, \bibinfo {author} {\bibfnamefont {M.}~\bibnamefont {Mignoli}}, \bibinfo {author} {\bibfnamefont {H.}~\bibnamefont {Pearce}}, \bibinfo {author} {\bibfnamefont {P.}~\bibnamefont {Rosati}}, \bibinfo {author} {\bibfnamefont {D.}~\bibnamefont {Stern}}, \bibinfo {author} {\bibfnamefont {L.}~\bibnamefont {Verde}}, \bibinfo {author} {\bibfnamefont {E.}~\bibnamefont {Zucca}}, \bibinfo {author} {\bibfnamefont {C.~M.}\ \bibnamefont {Carollo}}, \bibinfo {author} {\bibfnamefont {T.}~\bibnamefont
  {Contini}}, \bibinfo {author} {\bibfnamefont {J.-P.}\ \bibnamefont {Kneib}}, \bibinfo {author} {\bibfnamefont {O.}~\bibnamefont {Le~F{\`e}vre}}, \bibinfo {author} {\bibfnamefont {S.~J.}\ \bibnamefont {Lilly}}, \bibinfo {author} {\bibfnamefont {V.}~\bibnamefont {Mainieri}}, \emph {et~al.},\ }\bibfield  {title} {\bibinfo {title} {Improved constraints on the expansion rate of the universe up to z~1.1 from the spectroscopic evolution of cosmic chronometers},\ }\href {https://doi.org/10.1088/1475-7516/2012/08/006} {\bibfield  {journal} {\bibinfo  {journal} {Journal of Cosmology and Astroparticle Physics}\ }\textbf {\bibinfo {volume} {2012}}\bibfield  {number} {\bibinfo  {number} { (08)},\ \bibinfo {pages} {006}},\ }\Eprint {https://arxiv.org/abs/1201.3609} {arXiv:1201.3609 [astro-ph.CO]} \BibitemShut {NoStop}%
\bibitem [{\citenamefont {Loubser}(2025)}]{Loubser2025CC}%
  \BibitemOpen
  \bibfield  {author} {\bibinfo {author} {\bibfnamefont {S.~I.}\ \bibnamefont {Loubser}},\ }\href {https://arxiv.org/abs/2511.02730} {\bibinfo {title} {Measuring the expansion history of the universe with desi cosmic chronometers}} (\bibinfo {year} {2025}),\ \Eprint {https://arxiv.org/abs/2511.02730} {arXiv:2511.02730 [astro-ph.CO]} \BibitemShut {NoStop}%
\bibitem [{\citenamefont {Simon}\ \emph {et~al.}(2005)\citenamefont {Simon}, \citenamefont {Verde},\ and\ \citenamefont {Jimenez}}]{Simon2005}%
  \BibitemOpen
  \bibfield  {author} {\bibinfo {author} {\bibfnamefont {J.}~\bibnamefont {Simon}}, \bibinfo {author} {\bibfnamefont {L.}~\bibnamefont {Verde}},\ and\ \bibinfo {author} {\bibfnamefont {R.}~\bibnamefont {Jimenez}},\ }\bibfield  {title} {\bibinfo {title} {Constraints on the redshift dependence of the dark energy potential},\ }\href {https://doi.org/10.1103/PhysRevD.71.123001} {\bibfield  {journal} {\bibinfo  {journal} {Physical Review D}\ }\textbf {\bibinfo {volume} {71}},\ \bibinfo {pages} {123001} (\bibinfo {year} {2005})}\BibitemShut {NoStop}%
\bibitem [{\citenamefont {Jiao}\ \emph {et~al.}(2023)\citenamefont {Jiao}, \citenamefont {Borghi}, \citenamefont {Moresco},\ and\ \citenamefont {Zhang}}]{Jiao2023}%
  \BibitemOpen
  \bibfield  {author} {\bibinfo {author} {\bibfnamefont {K.}~\bibnamefont {Jiao}}, \bibinfo {author} {\bibfnamefont {N.}~\bibnamefont {Borghi}}, \bibinfo {author} {\bibfnamefont {M.}~\bibnamefont {Moresco}},\ and\ \bibinfo {author} {\bibfnamefont {T.-J.}\ \bibnamefont {Zhang}},\ }\bibfield  {title} {\bibinfo {title} {New observational h(z) data from full-spectrum fitting of cosmic chronometers in the lega-c survey},\ }\href {https://doi.org/10.3847/1538-4365/acbc77} {\bibfield  {journal} {\bibinfo  {journal} {Astrophys. J. Suppl. Ser.}\ }\textbf {\bibinfo {volume} {265}},\ \bibinfo {pages} {48} (\bibinfo {year} {2023})},\ \Eprint {https://arxiv.org/abs/2205.05701} {arXiv:2205.05701 [astro-ph.CO]} \BibitemShut {NoStop}%
\bibitem [{\citenamefont {Stern}\ \emph {et~al.}(2010)\citenamefont {Stern}, \citenamefont {Jimenez}, \citenamefont {Verde}, \citenamefont {Kamionkowski},\ and\ \citenamefont {Stanford}}]{Stern2010}%
  \BibitemOpen
  \bibfield  {author} {\bibinfo {author} {\bibfnamefont {D.}~\bibnamefont {Stern}}, \bibinfo {author} {\bibfnamefont {R.}~\bibnamefont {Jimenez}}, \bibinfo {author} {\bibfnamefont {L.}~\bibnamefont {Verde}}, \bibinfo {author} {\bibfnamefont {M.}~\bibnamefont {Kamionkowski}},\ and\ \bibinfo {author} {\bibfnamefont {S.~A.}\ \bibnamefont {Stanford}},\ }\bibfield  {title} {\bibinfo {title} {Cosmic chronometers: Constraining the equation of state of dark energy. i: H(z) measurements},\ }\href {https://doi.org/10.1088/1475-7516/2010/02/008} {\bibfield  {journal} {\bibinfo  {journal} {Journal of Cosmology and Astroparticle Physics}\ }\textbf {\bibinfo {volume} {2010}}\bibfield  {number} {\bibinfo  {number} { (02)},\ \bibinfo {pages} {008}},\ }\Eprint {https://arxiv.org/abs/0907.3149} {arXiv:0907.3149 [astro-ph.CO]} \BibitemShut {NoStop}%
\bibitem [{\citenamefont {Moresco}\ \emph {et~al.}(2016)\citenamefont {Moresco}, \citenamefont {Pozzetti}, \citenamefont {Cimatti}, \citenamefont {Jimenez}, \citenamefont {Maraston}, \citenamefont {Verde}, \citenamefont {Thomas}, \citenamefont {Citro}, \citenamefont {Tojeiro},\ and\ \citenamefont {Wilkinson}}]{Moresco2016}%
  \BibitemOpen
  \bibfield  {author} {\bibinfo {author} {\bibfnamefont {M.}~\bibnamefont {Moresco}}, \bibinfo {author} {\bibfnamefont {L.}~\bibnamefont {Pozzetti}}, \bibinfo {author} {\bibfnamefont {A.}~\bibnamefont {Cimatti}}, \bibinfo {author} {\bibfnamefont {R.}~\bibnamefont {Jimenez}}, \bibinfo {author} {\bibfnamefont {C.}~\bibnamefont {Maraston}}, \bibinfo {author} {\bibfnamefont {L.}~\bibnamefont {Verde}}, \bibinfo {author} {\bibfnamefont {D.}~\bibnamefont {Thomas}}, \bibinfo {author} {\bibfnamefont {A.}~\bibnamefont {Citro}}, \bibinfo {author} {\bibfnamefont {R.}~\bibnamefont {Tojeiro}},\ and\ \bibinfo {author} {\bibfnamefont {D.}~\bibnamefont {Wilkinson}},\ }\bibfield  {title} {\bibinfo {title} {A 6\% measurement of the hubble parameter at z~0.45: direct evidence of the epoch of cosmic re-acceleration},\ }\href {https://doi.org/10.1088/1475-7516/2016/05/014} {\bibfield  {journal} {\bibinfo  {journal} {Journal of Cosmology and Astroparticle Physics}\ }\textbf {\bibinfo {volume} {2016}}\bibfield  {number} {\bibinfo
  {number} { (05)},\ \bibinfo {pages} {014}},\ }\Eprint {https://arxiv.org/abs/1601.01701} {arXiv:1601.01701 [astro-ph.CO]} \BibitemShut {NoStop}%
\bibitem [{\citenamefont {Tomasetti}\ \emph {et~al.}(2023)\citenamefont {Tomasetti}, \citenamefont {Moresco}, \citenamefont {Borghi}, \citenamefont {Jiao}, \citenamefont {Cimatti}, \citenamefont {Pozzetti}, \citenamefont {Carnall}, \citenamefont {McLure},\ and\ \citenamefont {Pentericci}}]{Tomasetti2023}%
  \BibitemOpen
  \bibfield  {author} {\bibinfo {author} {\bibfnamefont {E.}~\bibnamefont {Tomasetti}}, \bibinfo {author} {\bibfnamefont {M.}~\bibnamefont {Moresco}}, \bibinfo {author} {\bibfnamefont {N.}~\bibnamefont {Borghi}}, \bibinfo {author} {\bibfnamefont {K.}~\bibnamefont {Jiao}}, \bibinfo {author} {\bibfnamefont {A.}~\bibnamefont {Cimatti}}, \bibinfo {author} {\bibfnamefont {L.}~\bibnamefont {Pozzetti}}, \bibinfo {author} {\bibfnamefont {A.~C.}\ \bibnamefont {Carnall}}, \bibinfo {author} {\bibfnamefont {R.~J.}\ \bibnamefont {McLure}},\ and\ \bibinfo {author} {\bibfnamefont {L.}~\bibnamefont {Pentericci}},\ }\bibfield  {title} {\bibinfo {title} {A new measurement of the expansion history of the universe at $z = 1.26$ with cosmic chronometers in vandels},\ }\href {https://doi.org/10.1051/0004-6361/202346992} {\bibfield  {journal} {\bibinfo  {journal} {Astronomy \& Astrophysics}\ }\textbf {\bibinfo {volume} {679}},\ \bibinfo {pages} {A96} (\bibinfo {year} {2023})}\BibitemShut {NoStop}%
\bibitem [{\citenamefont {Moresco}(2015)}]{Moresco2015}%
  \BibitemOpen
  \bibfield  {author} {\bibinfo {author} {\bibfnamefont {M.}~\bibnamefont {Moresco}},\ }\bibfield  {title} {\bibinfo {title} {Raising the bar: new constraints on the hubble parameter with cosmic chronometers at z $\sim$ 2},\ }\href {https://doi.org/10.1093/mnrasl/slv037} {\bibfield  {journal} {\bibinfo  {journal} {Mon. Not. R. Astron. Soc. Lett.}\ }\textbf {\bibinfo {volume} {450}},\ \bibinfo {pages} {L16} (\bibinfo {year} {2015})},\ \Eprint {https://arxiv.org/abs/1503.01116} {arXiv:1503.01116 [astro-ph.CO]} \BibitemShut {NoStop}%
\bibitem [{\citenamefont {Ratsimbazafy}\ \emph {et~al.}(2017)\citenamefont {Ratsimbazafy}, \citenamefont {Loubser}, \citenamefont {Crawford}, \citenamefont {Cress}, \citenamefont {Bassett}, \citenamefont {Nichol},\ and\ \citenamefont {V{\"a}is{\"a}nen}}]{Ratsimbazafy2017}%
  \BibitemOpen
  \bibfield  {author} {\bibinfo {author} {\bibfnamefont {A.~L.}\ \bibnamefont {Ratsimbazafy}}, \bibinfo {author} {\bibfnamefont {S.~I.}\ \bibnamefont {Loubser}}, \bibinfo {author} {\bibfnamefont {S.~M.}\ \bibnamefont {Crawford}}, \bibinfo {author} {\bibfnamefont {C.~M.}\ \bibnamefont {Cress}}, \bibinfo {author} {\bibfnamefont {B.~A.}\ \bibnamefont {Bassett}}, \bibinfo {author} {\bibfnamefont {R.~C.}\ \bibnamefont {Nichol}},\ and\ \bibinfo {author} {\bibfnamefont {P.}~\bibnamefont {V{\"a}is{\"a}nen}},\ }\bibfield  {title} {\bibinfo {title} {Age-dating luminous red galaxies observed with the southern african large telescope},\ }\href {https://doi.org/10.1093/mnras/stx301} {\bibfield  {journal} {\bibinfo  {journal} {Monthly Notices of the Royal Astronomical Society}\ }\textbf {\bibinfo {volume} {467}},\ \bibinfo {pages} {3239} (\bibinfo {year} {2017})},\ \Eprint {https://arxiv.org/abs/1702.00418} {arXiv:1702.00418 [astro-ph.CO]} \BibitemShut {NoStop}%
\bibitem [{\citenamefont {Peebles}(1993)}]{Peebles1993}%
  \BibitemOpen
  \bibfield  {author} {\bibinfo {author} {\bibfnamefont {P.~J.~E.}\ \bibnamefont {Peebles}},\ }\href@noop {} {\emph {\bibinfo {title} {Principles of Physical Cosmology}}}\ (\bibinfo  {publisher} {Princeton University Press},\ \bibinfo {address} {Princeton, NJ},\ \bibinfo {year} {1993})\BibitemShut {NoStop}%
\bibitem [{\citenamefont {Chevallier}\ and\ \citenamefont {Polarski}(2001)}]{Chevallier2001}%
  \BibitemOpen
  \bibfield  {author} {\bibinfo {author} {\bibfnamefont {M.}~\bibnamefont {Chevallier}}\ and\ \bibinfo {author} {\bibfnamefont {D.}~\bibnamefont {Polarski}},\ }\bibfield  {title} {\bibinfo {title} {Accelerating universes with scaling dark matter},\ }\href {https://doi.org/10.1142/S0218271801000822} {\bibfield  {journal} {\bibinfo  {journal} {International Journal of Modern Physics D}\ }\textbf {\bibinfo {volume} {10}},\ \bibinfo {pages} {213} (\bibinfo {year} {2001})},\ \Eprint {https://arxiv.org/abs/gr-qc/0009008} {arXiv:gr-qc/0009008} \BibitemShut {NoStop}%
\bibitem [{\citenamefont {Linder}(2003)}]{Linder2003}%
  \BibitemOpen
  \bibfield  {author} {\bibinfo {author} {\bibfnamefont {E.~V.}\ \bibnamefont {Linder}},\ }\bibfield  {title} {\bibinfo {title} {Exploring the expansion history of the universe},\ }\href {https://doi.org/10.1103/PhysRevLett.90.091301} {\bibfield  {journal} {\bibinfo  {journal} {Physical Review Letters}\ }\textbf {\bibinfo {volume} {90}},\ \bibinfo {pages} {091301} (\bibinfo {year} {2003})},\ \Eprint {https://arxiv.org/abs/astro-ph/0208512} {arXiv:astro-ph/0208512} \BibitemShut {NoStop}%
\bibitem [{\citenamefont {Visser}(2005)}]{Visser2005}%
  \BibitemOpen
  \bibfield  {author} {\bibinfo {author} {\bibfnamefont {M.}~\bibnamefont {Visser}},\ }\bibfield  {title} {\bibinfo {title} {Cosmography: Cosmology without the einstein equations},\ }\href {https://doi.org/10.1007/s10714-005-0134-8} {\bibfield  {journal} {\bibinfo  {journal} {Gen. Relativ. Gravit.}\ }\textbf {\bibinfo {volume} {37}},\ \bibinfo {pages} {1541} (\bibinfo {year} {2005})},\ \Eprint {https://arxiv.org/abs/gr-qc/0411131} {arXiv:gr-qc/0411131 [gr-qc]} \BibitemShut {NoStop}%
\bibitem [{\citenamefont {Gruber}\ and\ \citenamefont {Luongo}(2014)}]{Gruber2014}%
  \BibitemOpen
  \bibfield  {author} {\bibinfo {author} {\bibfnamefont {C.}~\bibnamefont {Gruber}}\ and\ \bibinfo {author} {\bibfnamefont {O.}~\bibnamefont {Luongo}},\ }\bibfield  {title} {\bibinfo {title} {Cosmographic analysis of the equation of state of the universe through pade approximations},\ }\href {https://doi.org/10.1103/PhysRevD.89.103506} {\bibfield  {journal} {\bibinfo  {journal} {Phys. Rev. D}\ }\textbf {\bibinfo {volume} {89}},\ \bibinfo {pages} {103506} (\bibinfo {year} {2014})},\ \Eprint {https://arxiv.org/abs/1309.3215} {arXiv:1309.3215 [gr-qc]} \BibitemShut {NoStop}%
\bibitem [{\citenamefont {Zhou}\ \emph {et~al.}(2016)\citenamefont {Zhou}, \citenamefont {Liu}, \citenamefont {Zou},\ and\ \citenamefont {Wei}}]{Zhou2016}%
  \BibitemOpen
  \bibfield  {author} {\bibinfo {author} {\bibfnamefont {Y.-N.}\ \bibnamefont {Zhou}}, \bibinfo {author} {\bibfnamefont {D.-Z.}\ \bibnamefont {Liu}}, \bibinfo {author} {\bibfnamefont {X.-B.}\ \bibnamefont {Zou}},\ and\ \bibinfo {author} {\bibfnamefont {H.}~\bibnamefont {Wei}},\ }\bibfield  {title} {\bibinfo {title} {New generalizations of cosmography inspired by the pad{\'e} approximant},\ }\href {https://doi.org/10.1140/epjc/s10052-016-4091-z} {\bibfield  {journal} {\bibinfo  {journal} {Eur. Phys. J. C}\ }\textbf {\bibinfo {volume} {76}},\ \bibinfo {pages} {281} (\bibinfo {year} {2016})},\ \Eprint {https://arxiv.org/abs/1602.07189} {arXiv:1602.07189 [gr-qc]} \BibitemShut {NoStop}%
\bibitem [{\citenamefont {Capozziello}\ \emph {et~al.}(2019{\natexlab{a}})\citenamefont {Capozziello}, \citenamefont {Ruchika},\ and\ \citenamefont {Sen}}]{Capozziello2019a}%
  \BibitemOpen
  \bibfield  {author} {\bibinfo {author} {\bibfnamefont {S.}~\bibnamefont {Capozziello}}, \bibinfo {author} {\bibnamefont {Ruchika}},\ and\ \bibinfo {author} {\bibfnamefont {A.~A.}\ \bibnamefont {Sen}},\ }\bibfield  {title} {\bibinfo {title} {Model-independent constraints on dark energy evolution from low-redshift observations},\ }\href {https://doi.org/10.1093/mnras/stz176} {\bibfield  {journal} {\bibinfo  {journal} {Mon. Not. R. Astron. Soc.}\ }\textbf {\bibinfo {volume} {484}},\ \bibinfo {pages} {4484} (\bibinfo {year} {2019}{\natexlab{a}})},\ \Eprint {https://arxiv.org/abs/1806.03943} {arXiv:1806.03943 [astro-ph.CO]} \BibitemShut {NoStop}%
\bibitem [{\citenamefont {Capozziello}\ \emph {et~al.}(2019{\natexlab{b}})\citenamefont {Capozziello}, \citenamefont {D'Agostino},\ and\ \citenamefont {Luongo}}]{Capozziello2019b}%
  \BibitemOpen
  \bibfield  {author} {\bibinfo {author} {\bibfnamefont {S.}~\bibnamefont {Capozziello}}, \bibinfo {author} {\bibfnamefont {R.}~\bibnamefont {D'Agostino}},\ and\ \bibinfo {author} {\bibfnamefont {O.}~\bibnamefont {Luongo}},\ }\bibfield  {title} {\bibinfo {title} {Extended gravity cosmography},\ }\href {https://doi.org/10.1142/S0218271819300167} {\bibfield  {journal} {\bibinfo  {journal} {Int. J. Mod. Phys. D}\ }\textbf {\bibinfo {volume} {28}},\ \bibinfo {pages} {1930016} (\bibinfo {year} {2019}{\natexlab{b}})},\ \Eprint {https://arxiv.org/abs/1904.01427} {arXiv:1904.01427 [gr-qc]} \BibitemShut {NoStop}%
\bibitem [{\citenamefont {Aviles}\ \emph {et~al.}(2014)\citenamefont {Aviles}, \citenamefont {Bravetti}, \citenamefont {Capozziello},\ and\ \citenamefont {Luongo}}]{Aviles2014}%
  \BibitemOpen
  \bibfield  {author} {\bibinfo {author} {\bibfnamefont {A.}~\bibnamefont {Aviles}}, \bibinfo {author} {\bibfnamefont {A.}~\bibnamefont {Bravetti}}, \bibinfo {author} {\bibfnamefont {S.}~\bibnamefont {Capozziello}},\ and\ \bibinfo {author} {\bibfnamefont {O.}~\bibnamefont {Luongo}},\ }\bibfield  {title} {\bibinfo {title} {Precision cosmology with pade rational approximations: Theoretical predictions versus observational limits},\ }\href {https://doi.org/10.1103/PhysRevD.90.043531} {\bibfield  {journal} {\bibinfo  {journal} {Phys. Rev. D}\ }\textbf {\bibinfo {volume} {90}},\ \bibinfo {pages} {043531} (\bibinfo {year} {2014})},\ \Eprint {https://arxiv.org/abs/1405.6935} {arXiv:1405.6935 [gr-qc]} \BibitemShut {NoStop}%
\bibitem [{\citenamefont {Hu}\ \emph {et~al.}(2024)\citenamefont {Hu}, \citenamefont {Hu}, \citenamefont {Jia}, \citenamefont {Gao},\ and\ \citenamefont {Wang}}]{Hu2024}%
  \BibitemOpen
  \bibfield  {author} {\bibinfo {author} {\bibfnamefont {J.~P.}\ \bibnamefont {Hu}}, \bibinfo {author} {\bibfnamefont {J.}~\bibnamefont {Hu}}, \bibinfo {author} {\bibfnamefont {X.~D.}\ \bibnamefont {Jia}}, \bibinfo {author} {\bibfnamefont {B.~Q.}\ \bibnamefont {Gao}},\ and\ \bibinfo {author} {\bibfnamefont {F.~Y.}\ \bibnamefont {Wang}},\ }\bibfield  {title} {\bibinfo {title} {Testing cosmic anisotropy with pade approximations and the latest pantheon+ sample},\ }\href {https://doi.org/10.1051/0004-6361/202450342} {\bibfield  {journal} {\bibinfo  {journal} {Astron. Astrophys.}\ }\textbf {\bibinfo {volume} {689}},\ \bibinfo {pages} {A215} (\bibinfo {year} {2024})},\ \Eprint {https://arxiv.org/abs/2406.14827} {arXiv:2406.14827 [astro-ph.CO]} \BibitemShut {NoStop}%
\bibitem [{\citenamefont {Pourojaghi}\ \emph {et~al.}(2025)\citenamefont {Pourojaghi}, \citenamefont {Malekjani},\ and\ \citenamefont {Davari}}]{Pourojaghi2025}%
  \BibitemOpen
  \bibfield  {author} {\bibinfo {author} {\bibfnamefont {S.}~\bibnamefont {Pourojaghi}}, \bibinfo {author} {\bibfnamefont {M.}~\bibnamefont {Malekjani}},\ and\ \bibinfo {author} {\bibfnamefont {Z.}~\bibnamefont {Davari}},\ }\bibfield  {title} {\bibinfo {title} {{$\Lambda$CDM model against cosmography: a possible deviation after DESI 2024}},\ }\href {https://doi.org/10.1093/mnras/staf037} {\bibfield  {journal} {\bibinfo  {journal} {Monthly Notices of the Royal Astronomical Society}\ }\textbf {\bibinfo {volume} {537}},\ \bibinfo {pages} {436} (\bibinfo {year} {2025})},\ \Eprint {https://arxiv.org/abs/2408.10704} {arXiv:2408.10704 [astro-ph.CO]} \BibitemShut {NoStop}%
\bibitem [{\citenamefont {Petreca}\ \emph {et~al.}(2024)\citenamefont {Petreca}, \citenamefont {Benetti},\ and\ \citenamefont {Capozziello}}]{Petreca2024}%
  \BibitemOpen
  \bibfield  {author} {\bibinfo {author} {\bibfnamefont {A.~T.}\ \bibnamefont {Petreca}}, \bibinfo {author} {\bibfnamefont {M.}~\bibnamefont {Benetti}},\ and\ \bibinfo {author} {\bibfnamefont {S.}~\bibnamefont {Capozziello}},\ }\bibfield  {title} {\bibinfo {title} {Beyond $\lambda$cdm with $f(z)$cdm: Criticalities and solutions of pad\'e cosmography},\ }\href {https://doi.org/10.1016/j.dark.2024.101453} {\bibfield  {journal} {\bibinfo  {journal} {Physics of the Dark Universe}\ }\textbf {\bibinfo {volume} {44}},\ \bibinfo {pages} {101453} (\bibinfo {year} {2024})}\BibitemShut {NoStop}%
\bibitem [{\citenamefont {Visser}(2004)}]{visser2004}%
  \BibitemOpen
  \bibfield  {author} {\bibinfo {author} {\bibfnamefont {M.}~\bibnamefont {Visser}},\ }\bibfield  {title} {\bibinfo {title} {Jerk, snap and the cosmological equation of state},\ }\href {https://doi.org/10.1088/0264-9381/21/11/006} {\bibfield  {journal} {\bibinfo  {journal} {Classical and Quantum Gravity}\ }\textbf {\bibinfo {volume} {21}},\ \bibinfo {pages} {2603} (\bibinfo {year} {2004})}\BibitemShut {NoStop}%
\bibitem [{\citenamefont {Wang}\ \emph {et~al.}(2009)\citenamefont {Wang}, \citenamefont {Dai},\ and\ \citenamefont {Qi}}]{Wang2009}%
  \BibitemOpen
  \bibfield  {author} {\bibinfo {author} {\bibfnamefont {F.~Y.}\ \bibnamefont {Wang}}, \bibinfo {author} {\bibfnamefont {Z.~G.}\ \bibnamefont {Dai}},\ and\ \bibinfo {author} {\bibfnamefont {S.}~\bibnamefont {Qi}},\ }\bibfield  {title} {\bibinfo {title} {Probing the cosmographic parameters to distinguish between dark energy and modified gravity models},\ }\href {https://doi.org/10.1051/0004-6361/200911998} {\bibfield  {journal} {\bibinfo  {journal} {Astronomy \& Astrophysics}\ }\textbf {\bibinfo {volume} {507}},\ \bibinfo {pages} {53} (\bibinfo {year} {2009})}\BibitemShut {NoStop}%
\bibitem [{\citenamefont {Foreman-Mackey}\ \emph {et~al.}(2013)\citenamefont {Foreman-Mackey}, \citenamefont {Hogg}, \citenamefont {Lang},\ and\ \citenamefont {Goodman}}]{ForemanMackey2013}%
  \BibitemOpen
  \bibfield  {author} {\bibinfo {author} {\bibfnamefont {D.}~\bibnamefont {Foreman-Mackey}}, \bibinfo {author} {\bibfnamefont {D.~W.}\ \bibnamefont {Hogg}}, \bibinfo {author} {\bibfnamefont {D.}~\bibnamefont {Lang}},\ and\ \bibinfo {author} {\bibfnamefont {J.}~\bibnamefont {Goodman}},\ }\bibfield  {title} {\bibinfo {title} {emcee: The mcmc hammer},\ }\href {https://doi.org/10.1086/670067} {\bibfield  {journal} {\bibinfo  {journal} {Publications of the Astronomical Society of the Pacific}\ }\textbf {\bibinfo {volume} {125}},\ \bibinfo {pages} {306} (\bibinfo {year} {2013})}\BibitemShut {NoStop}%
\bibitem [{\citenamefont {Tamayo}\ and\ \citenamefont {V{\'a}zquez}(2019)}]{Tamayo2019}%
  \BibitemOpen
  \bibfield  {author} {\bibinfo {author} {\bibfnamefont {D.}~\bibnamefont {Tamayo}}\ and\ \bibinfo {author} {\bibfnamefont {J.~A.}\ \bibnamefont {V{\'a}zquez}},\ }\bibfield  {title} {\bibinfo {title} {Fourier-series expansion of the dark-energy equation of state},\ }\href {https://doi.org/10.1093/mnras/stz1229} {\bibfield  {journal} {\bibinfo  {journal} {Monthly Notices of the Royal Astronomical Society}\ }\textbf {\bibinfo {volume} {487}},\ \bibinfo {pages} {729} (\bibinfo {year} {2019})}\BibitemShut {NoStop}%
\bibitem [{\citenamefont {Dainotti}\ \emph {et~al.}(2021)\citenamefont {Dainotti}, \citenamefont {De~Simone}, \citenamefont {Schiavone}, \citenamefont {Montani}, \citenamefont {Rinaldi},\ and\ \citenamefont {Lambiase}}]{Dainotti2021}%
  \BibitemOpen
  \bibfield  {author} {\bibinfo {author} {\bibfnamefont {M.~G.}\ \bibnamefont {Dainotti}}, \bibinfo {author} {\bibfnamefont {B.}~\bibnamefont {De~Simone}}, \bibinfo {author} {\bibfnamefont {T.}~\bibnamefont {Schiavone}}, \bibinfo {author} {\bibfnamefont {G.}~\bibnamefont {Montani}}, \bibinfo {author} {\bibfnamefont {E.}~\bibnamefont {Rinaldi}},\ and\ \bibinfo {author} {\bibfnamefont {G.}~\bibnamefont {Lambiase}},\ }\bibfield  {title} {\bibinfo {title} {On the hubble constant tension in the sne ia pantheon sample},\ }\href {https://doi.org/10.3847/1538-4357/abeb73} {\bibfield  {journal} {\bibinfo  {journal} {The Astrophysical Journal}\ }\textbf {\bibinfo {volume} {912}},\ \bibinfo {pages} {150} (\bibinfo {year} {2021})}\BibitemShut {NoStop}%
\bibitem [{\citenamefont {Montani}\ \emph {et~al.}(2025)\citenamefont {Montani}, \citenamefont {Carlevaro},\ and\ \citenamefont {Dainotti}}]{Montani2025}%
  \BibitemOpen
  \bibfield  {author} {\bibinfo {author} {\bibfnamefont {G.}~\bibnamefont {Montani}}, \bibinfo {author} {\bibfnamefont {N.}~\bibnamefont {Carlevaro}},\ and\ \bibinfo {author} {\bibfnamefont {M.~G.}\ \bibnamefont {Dainotti}},\ }\bibfield  {title} {\bibinfo {title} {Running hubble constant: Evolutionary dark energy},\ }\href {https://doi.org/10.1016/j.dark.2025.101847} {\bibfield  {journal} {\bibinfo  {journal} {Physics of the Dark Universe}\ }\textbf {\bibinfo {volume} {48}},\ \bibinfo {pages} {101847} (\bibinfo {year} {2025})}\BibitemShut {NoStop}%
\bibitem [{\citenamefont {Mo}(2026)}]{Mo2026Zenodo}%
  \BibitemOpen
  \bibfield  {author} {\bibinfo {author} {\bibfnamefont {Z.-Y.}\ \bibnamefont {Mo}},\ }\href {https://doi.org/10.5281/zenodo.18345843} {\bibinfo {title} {Data and code for ``redshift-binned constraints on the hubble constant under {Lambda}cdm, {CPL}, and {Pad\'e} cosmography''}},\ \bibinfo {howpublished} {(v1.0.0) (Zenodo, 2026)} (\bibinfo {year} {2026})\BibitemShut {NoStop}%
\end{thebibliography}%

\end{document}